%% file: main-PLB.tex
\documentclass[preprint,5p,times,twocolumn]{elsarticle}
\biboptions{sort&compress}

\usepackage{amssymb}
\usepackage{lipsum}

\journal{Physics Letters B}

\input{macro}

\begin{document}

\begin{frontmatter}

\title{Hyperon global polarization in isobar Ru+Ru and Zr+Zr collisions at \snn = 200 GeV}

\author{The STAR Collaboration}

\begin{abstract}
The polarization of $\Lambda$, $\bar{\Lambda}$, $\Xi^-$, and $\bar{\Xi}^+$ hyperons along the angular momentum of the system has been measured in isobar collisions of Ru+Ru and Zr+Zr at \snn = 200 GeV with the STAR detector at RHIC. 
The polarization dependence on collision centrality is explored and found to show an increasing trend in more peripheral collisions. 
Dependencies on transverse momentum and pseudorapidity are investigated for $\Lambda$ and $\bar{\Lambda}$ hyperons, but no significant dependence has been observed. 
The polarization measurements for $\Lambda$ and $\bar{\Lambda}$ are consistent with each other, indicating little contribution of the spin-magnetic coupling in the observed polarization. 
The results for $\Lambda$ hyperons measurements are qualitatively consistent with hydrodynamic calculations incorporating effects from shear-induced polarization and thermal vorticity, and show no obvious system size dependence in comparison with previous results in Au+Au collisions. 
For the first time, the dependence of the polarization on the hyperon's emission azimuthal angle with respect to the second harmonic event plane is extracted and shows stronger polarization for the in-plane emitted hyperons at the level of 2.4$\sigma$ significance in 20-50\% centrality. 
The measurements of $\Xi$ hyperons polarization via the polarization transfer analysis exhibit a finite positive polarization, 2.9$\sigma$ significance in 20-50\% centrality, slightly enhanced compared to the inclusive $\Lambda$ polarization.
\end{abstract}

\begin{keyword}
global polarization \sep hyperon \sep vorticity

\PACS 25.75.-q \sep 25.75.Ld \sep 24.70.+s
\end{keyword}

\end{frontmatter}




\section{Introduction}
\label{introduction}

Spin, as a fundamental property of particles, has played a very important role in the development of different fields of physics, such as spintronics, shell structure in nuclear physics, and many others. 
In recent years, spin-related studies have also drawn great attention in high-energy nuclear physics. This interest was triggered particularly after the global polarization of hyperons and vector meson spin alignment in non-central nucleus-nucleus collisions~\cite{1,Voloshin:2004ha,Liang:2004xn,2,Gao:2007bc} were observed by experimental measurements in heavy-ion collisions~\cite{STAR:2017ckg,STAR:2022fan}.
The global polarization is a phenomenon referring to non-zero spin particles being polarized on average along the initial orbital angular momentum of the system, due to spin-orbit coupling.
Since the first observations, the study of the polarization effects in nucleus-nucleus collisions has opened up new avenues for exploring the properties of nuclear matter.

Measurements of the global polarization of \lam hyperons have been extended to heavier hyperons such as $\Xi$ and $\Omega$~\cite{STAR:2020xbm}, and to collision energies from a few GeVs up to TeVs~\cite{STAR:2007ccu,STAR:2018gyt,ALICE:2019onw,STAR:2021beb,STAR:2023nvo,HADES:2022enx}.
The energy dependence of the global polarization can be well described by theoretical models based on thermal vorticity~\cite{Deng:2016gyh,Wei:2018zfb,Guo:2021udq,Bozek:2010bi,Ivanov:2018eej,Wu:2019eyi,Ivanov:2019ern,Ivanov:2020wak,Liang:2019pst,Jiang:2016woz}.  
Despite the significant progress, additional measurements are needed to probe unexplored effects and answer remaining questions, such as the splitting of the \lam and \alam global polarization in a strong magnetic field induced by the two charged ions at nearly speed of light~\cite{McLerran:2013hla,Becattini:2016gvu,Muller:2018ibh}, and testing the predicted rapidity and system size dependencies~\cite{Wu:2019eyi,Xie:2019jun,Liang:2019pst,Shi:2017wpk,Alzhrani:2022dpi}. 
The observed azimuthal angle dependence of the hyperon polarization along the beam direction~\cite{STAR:2019erd,ALICE:2021pzu,STAR:2023eck} also remains a puzzle that needs to be unraveled (see recent reviews for more details~\cite{Becattini:2020ngo,Becattini:2024uha,Niida:2024ntm,Chen:2024aom}).

Data from the isobar collisions, $^{96}_{44}$Ru+$^{96}_{44}$Ru and $^{96}_{40}$Zr+$^{96}_{40}$Zr, at \snn = 200 GeV have been collected by the STAR experiment for the study of the chiral magnetic effect~\cite{STAR:2021mii}.  
These species were chosen to allow for an difference in electric charge while keeping the total number of nucleons in each nucleus the same. 
As a result the square of the initial magnetic field is expected to be $\sim$15\% higher in Ru+Ru than in Zr+Zr collisions~\cite{Voloshin:2010ut,Deng:2016knn}.
These high-statistics isobar collision data can be a good tool to explore a possible effect of spin-magnetic coupling, 
which manifest as a difference in the global polarization of particles and antiparticles, owing to the opposite sign of their magnetic moments. 
Furthermore, theoretical model predicts larger polarization in smaller systems. This is because of the shorter lifetime of smaller system since the system vorticity is expected to be diluted with time~\cite{Shi:2017wpk,Alzhrani:2022dpi}. Thus the isobar collisions also provide an opportunity to study the possible system size dependence of the polarization by comparing to previous results in Au+Au collisions.

In this letter, we report precision measurements of $\Lambda$ and $\bar\Lambda$ global polarization in Ru+Ru and Zr+Zr collisions at \snn = 200 GeV with data collected by the STAR experiment at RHIC. 
We present differential measurements, including the dependence of the polarization on transverse momentum, pseudorapidity, and for the first time on azimuthal angle of hyperons.
In addition, we also report the global polarization measurements of $\Xi^-$ and $\bar \Xi^+$ hyperons.

\section{STAR Experiment}

The subsystems of the STAR detector used in this analysis are the time projection chamber (TPC)~\cite{Anderson:2003ur}, the time-of-flight (TOF) detector~\cite{Llope:2012zz},  the zero degree calorimeters (ZDCs)~\cite{Adler:2001fq}, and vertex position detectors (VPDs)~\cite{Llope:2014nva}. 
The TPC is the main tracking detector of the STAR experiment and is capable of measuring charged particles within the pseudorapidity range of $|\eta|<1$, and with full azimuthal coverage. 
The momentum of charged tracks can be determined based on their curvature in a constant magnetic field (0.5 Tesla) inside the TPC. The TPC also measures the specific ionization energy loss $dE/dx$ of the charged particles~\cite{Anderson:2003ur} which can be used for particle identification. 
The TOF detector extends the capability of particle identification up to the tranverse momentum $p_T$ = 3 GeV/$c$, located outside the TPC covering the full azimuth and a pseudorapidity range of $|\eta| < 0.9$~\cite{Llope:2012zz}. 
The ZDCs are forward/backward ($|\eta| > 6.3$) detectors that are used to measure the spectator neutrons~\cite{Adler:2001fq}. 
They are positioned 18 meters from the center of the interaction region, downstream of the return dipole magnets.
The installation of the Shower Maximum Detectors (SMDs)~\cite{SMD} with the ZDC capture the energy deposition and spatial transverse distribution of spectator neutrons. 
The VPDs are comprised of two identical detectors that encircle the beam pipe and cover \(4.24 < |\eta| < 5.1\), providing the information about the collision time and the collision vertex position along the beamline. 
The minimum-bias trigger used in this analysis is based on VPD coincidences.

\section{Data analysis}

This analysis is based on minimum-bias triggered events from Ru+Ru and Zr+Zr collisions at \snn = 200 GeV  collected in the year 2018. 
The collision vertices were reconstructed using charged-particle tracks measured in the TPC. 
Events were selected to have the collision vertex position within the asymmetric range of (-35, 25)~cm from the center of the TPC in the beam direction, and within $\pm$2~cm in the radial direction relative to the beam center. 
Due to online vertex selection, the vertex distribution was found to be asymmetric with a peak around $\text{-}5$ cm. 
Therefore the asymmetric cut was applied to maximize the statistics.
Pile-up events were suppressed by requiring the difference between the TPC and VPD vertices along the beam direction to be less than 5 cm, and also by excluding outliers in the correlation between the number of TPC tracks and the number of those tracks matched to hits in the TOF.
In total, 1.8 (2.0) billion minimum-bias events for Ru+Ru (Zr+Zr) collisions were used in the analysis. 
The centrality of each collision was determined by measuring event charged particle multiplicity and interpreting the measurement with a tuned Monte Carlo Glauber calculation~\cite{STAR:2021mii,Miller:2007ri}.

\subsection{Track selection}

Charged tracks reconstructed from the TPC hit information were selected based on the following criteria.  
The number of hit points used in the reconstruction was required to be greater than 15 to ensure good quality of tracks. 
The ratio of the number of hit points to the maximum possible number of hit points (45 at most depending on the track trajectory inside the TPC) was also required to be larger than 0.52 to avoid split tracks.
Charged tracks within the transverse momentum range of $0.15 < p_{\rm T} < 10$ GeV/$c$ and $|\eta| < 1$ were analyzed in this study.

\subsection{Hyperon reconstruction}

\begin{figure}[htb]
    \centering
    \hspace{-2mm}
    \subfigure
    {
    \includegraphics[width=0.49\linewidth]{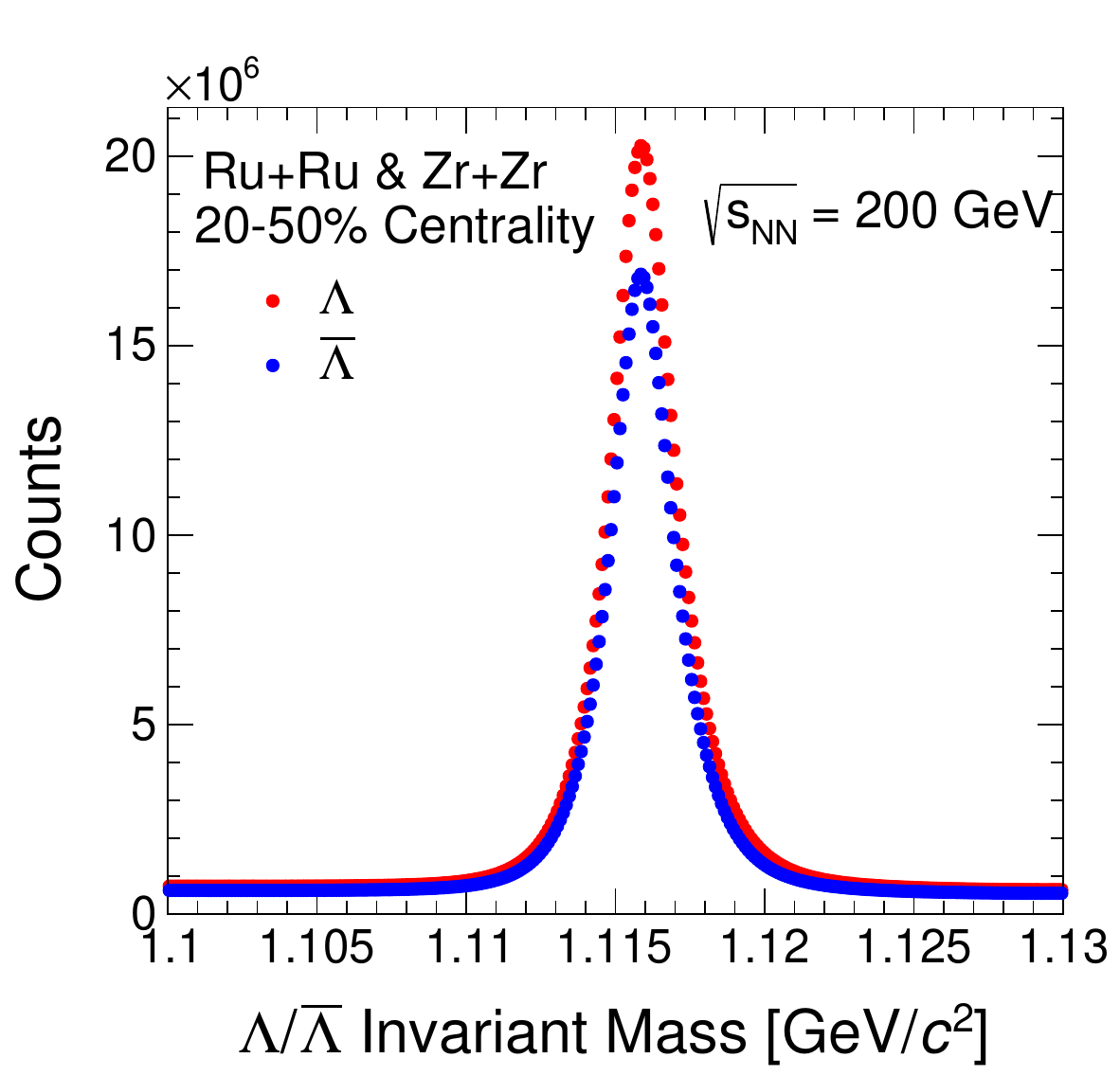}
    }\hspace{-3mm}
    \subfigure
    {
    \includegraphics[width=0.49\linewidth]{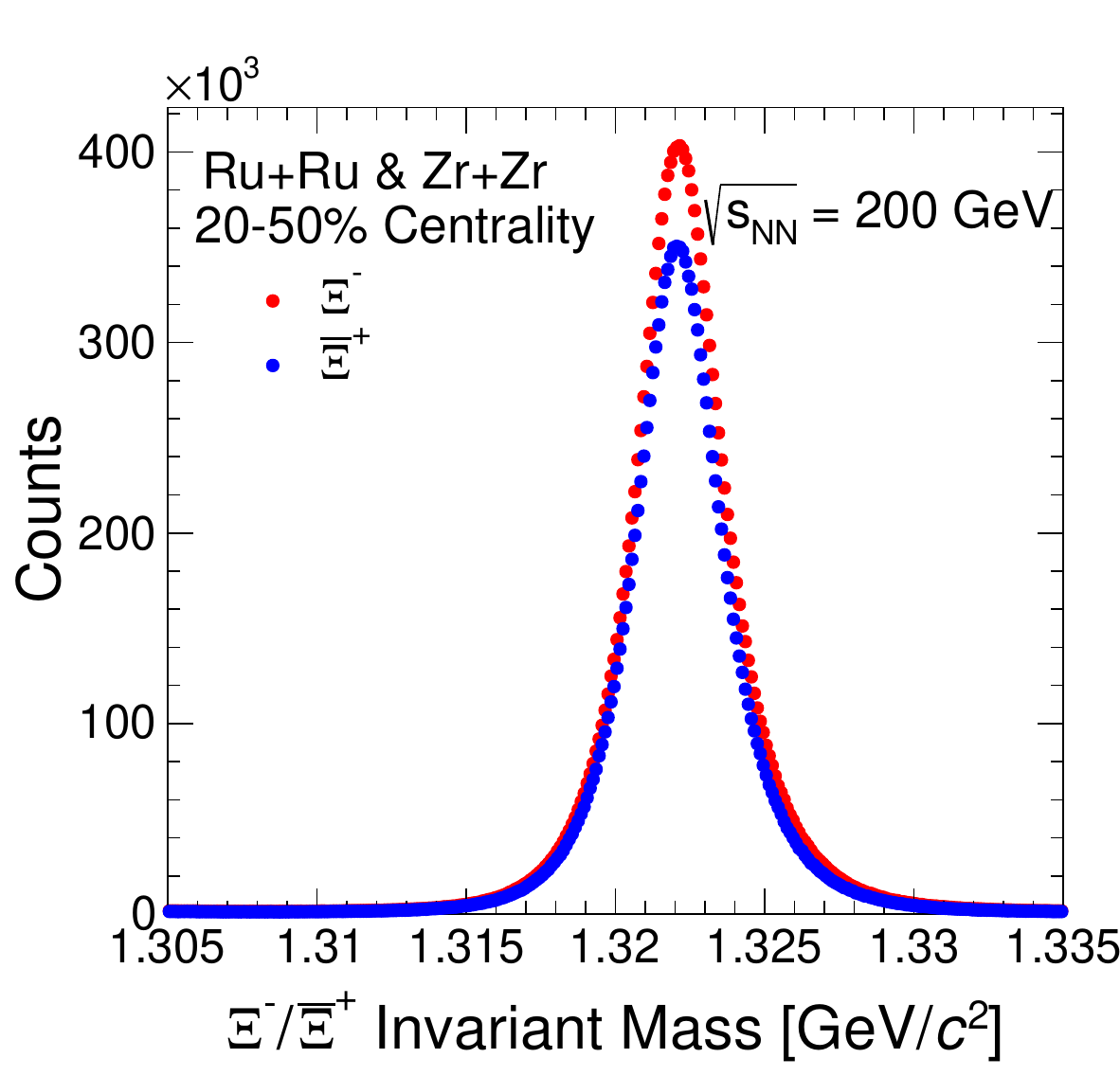}
    }\vspace{-3mm}
    \caption{Invariant mass distributions for $\Lambda$ and $\bar{\Lambda}$ (left) and for $\Xi^{-}$ and $\Bar{\Xi}^{+}$ (right) for 20-50\% centrality in Ru+Ru and Zr+Zr collisions at \snn = 200 GeV.}
    \label{fig:LambdaMass}
\end{figure}

$\Lambda$ and $\Bar{\Lambda}$ hyperons were reconstructed from their weak decay topology, $\Lambda \rightarrow p+\pi^{-}$ and $\Bar{\Lambda} \rightarrow \Bar{p} + \pi^{+}$. 
The parent $\Xi^{-}$/$\Bar{\Xi}^{+}$ and their daughter $\Lambda$/$\Bar{\Lambda}$ were reconstructed utilizing the decay channels of $\Xi^{-} \rightarrow \Lambda+\pi^{-}$ and $\Bar{\Xi}^{+} \rightarrow \Bar{\Lambda} + \pi^{+}$.
The identification of the $p$ ($\Bar{p}$ ) and $\pi^{-}$ ($\pi^{+}$) candidates is based on their electric charge and energy loss, $dE/dx$, in the TPC. 
Furthermore, the TOF detector was utilized if the track hits the TOF, which allows to identify the particles to higher $p_{\rm T}$.  
The invariant mass was calculated for all proton-pion pairs to find \lam candidates, and the optimized selection criteria based on the decay topology were applied to suppress combinatorial background.
These criteria include a maximum cut (0.85 cm) on the closest distance of approach (DCA) between the helical tracks of the proton and pion, a minimum requirement (0.35 cm for $p$, 1.1 cm for $\pi$) on the DCA between each track and the primary vertex, an upper limit (1.1 cm) on the DCA between the $\Lambda$ hyperon candidate and the primary vertex, and a lower threshold (4.0 cm) on the decay length of the $\Lambda$ hyperon. 
The reconstruction of $\Xi^{-}$($\bar\Xi^+$) hyperon was performed using the KF Particle Finder package, based on the Kalman Filter method initially developed for the CBM and ALICE experiments~\cite{Gorbunov:2013yvt,Zyzak:2016exl,Kisel:2018nvd}, to enhance $\Xi$ hyperon yield with a good purity.
This methodology leverages the track fit quality and decay topology to optimize the selection process.
Figure~\ref{fig:LambdaMass} shows the invariant mass distributions for $\Lambda$, $\Bar{\Lambda}$, $\Xi^{-}$, and $\Bar{\Xi}^{+}$ in the 20-50\% centrality bin as an example. 
The combinatorial background under the mass peak was estimated using off-peak regions and the background fraction is found to be smaller than 10\% (2\%) for \lam ($\Xi$) in 20-50\% centrality.

\subsection{Event plane determination}
 
In order to reconstruct the direction of the initial orbital angular momentum of the system, one needs to determine the azimuthal angle of the impact parameter vector, which is defined as a vector connecting the centers of the two nuclei and orthogonal to the system angular momentum direction.
The first-order event plane angle $\Psi_{1}$ as an experimental estimate of the impact parameter direction~\cite{Voloshin:2016ppr} was reconstructed by using the ZDC-SMDs, which determine the deflection angle of spectator neutrons in the transverse plane.
Also, to study azimuthal angle dependence of the polarization, the second-order event plane angle $\Psi_{2}$ was reconstructed based on the azimuthal distribution of charged tracks from the TPC within $0.2<p_{\rm T}<2$ GeV/$c$ and $|\eta|<1$.

\begin{figure}[htb]
    \centering
    \hspace{-2mm}
    \subfigure
    {
    \includegraphics[width=0.49\linewidth,clip]{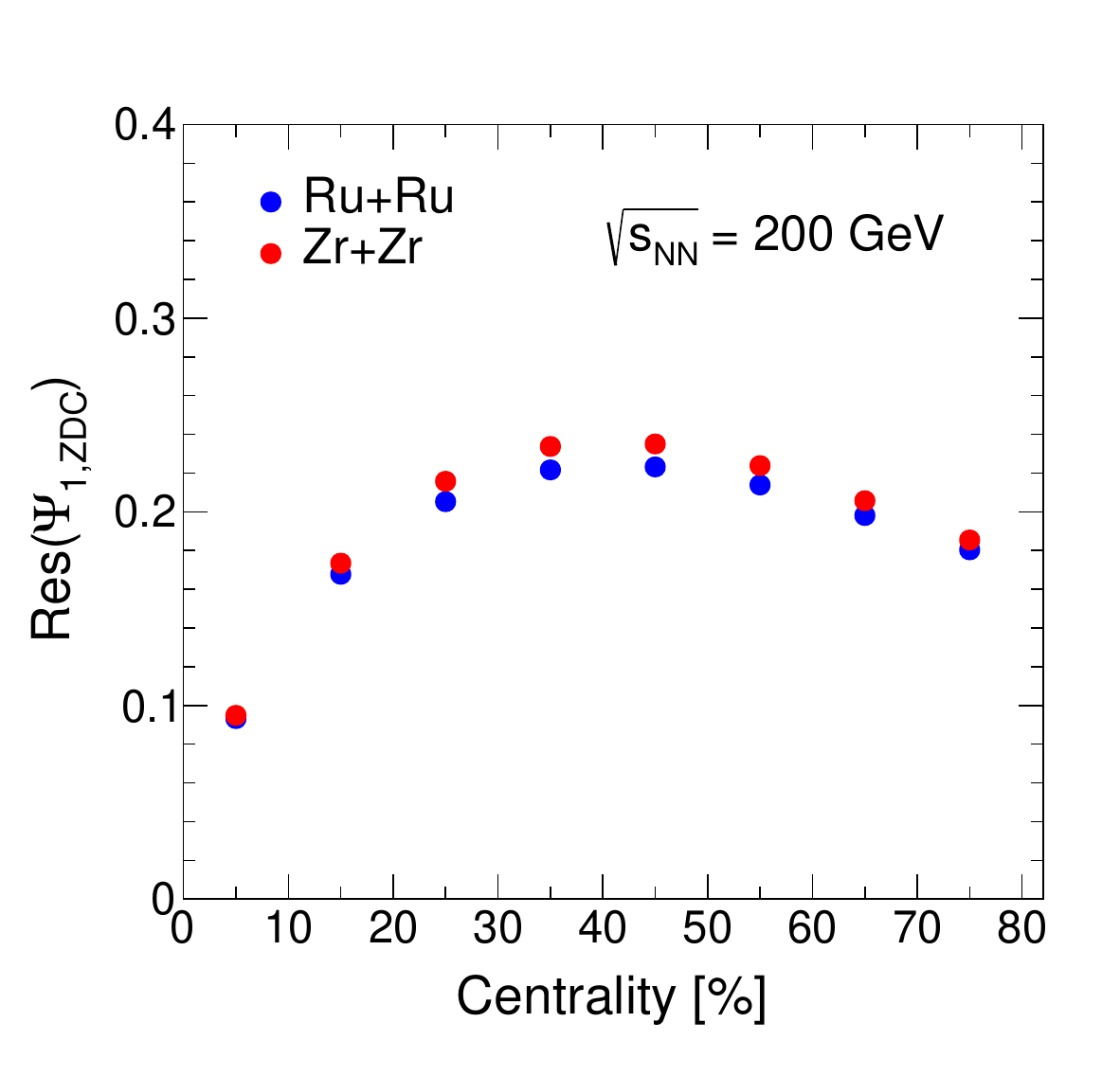}
    }\hspace{-3mm}
    \subfigure
    {
    \includegraphics[width=0.49\linewidth,clip]{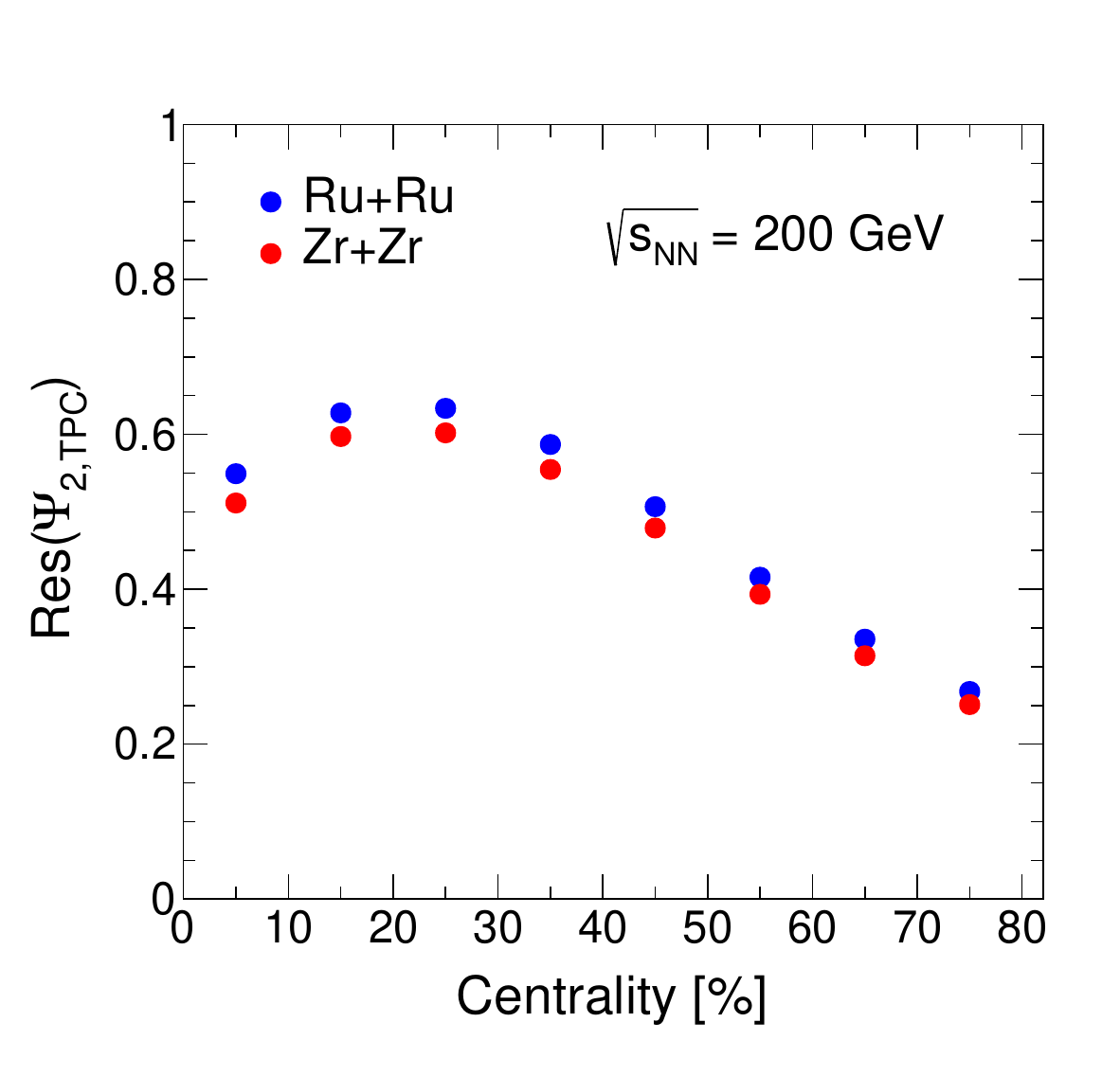}
    }\vspace{-3mm}
    \caption{Resolution of the first-order event plane from ZDC-SMDs (left) and the second-order event plane from TPC (right) as a function of collision centrality in Ru+Ru and Zr+Zr collisions at \snn = 200 GeV.}
    \label{fig:ZDCPsi1Resolution}
\end{figure}

The event plane resolution, Res($\Psi_{n}$)\,=\,$\langle \cos[n(\Psi_{n}^{\rm obs}-\Psi_{n})]\rangle$, was estimated by correlating two angles from forward and backward regions called the two-sub event method~\cite{Poskanzer:1998yz}, where $\Psi_{n}^{\rm obs}$ denotes a measured $n$th harmonic event plane. 
Figure~\ref{fig:ZDCPsi1Resolution} demonstrates the first- and second-order event plane resolutions as a function of collision centrality. 
The resolutions are very similar for the two isobar systems, with peaks at approximately 0.24 and 0.60 for the first- and second-order event planes, respectively. 
Note that the resolution for the first-order event plane ($\Psi_1$) is slightly larger in Zr+Zr collisions due to the greater number of neutrons contributing to $\Psi_1$. In contrast, the resolution for the second-order event plane ($\Psi_2$) is slightly larger in Ru+Ru collisions, which can be explained by the larger multiplicity of charged particles at midrapidity contributing to $\Psi_2$~\cite{STAR:2021mii}.

\subsection{Polarization measurement}
\label{Polarization_measurement}

The polarization along the initial angular momentum direction, perpendicular to the impact parameter vector, 
can be defined as~\cite{STAR:2007ccu},
\begin{equation}
    P_{\rm H} = \frac{8}{\pi\alpha_{\rm H}} \frac{1}{A_0} \frac{\langle\sin(\Psi_1-\phi_{\rm B}^{\ast})\rangle}{\rm Res(\Psi_1)}.
    \label{eq:PH}
\end{equation}
Here $P_{\rm H}$ is the polarization of hyperon, $\phi_{\rm B}^\ast$ is the azimuthal angle of the daughter baryon in the hyperon rest frame.
The decay parameters for $\Lambda$ and $\Xi$ hyperons are set to $\alpha_{\Lambda}$\,=\,$-\alpha_{\Bar{\Lambda}}$\,=\,0.732$\pm$0.014 and $\alpha_{\Xi^{-}}$\,=\,-$\alpha_{\Bar{\Xi}^{+}}$\,=\,-0.401$\pm$0.058~\cite{ParticleDataGroup:2022pth}, assuming the charge conjugation parity symmetry.
The acceptance correction factor $A_0$ is estimated from the data (see Eq.~\eqref{eq:A0}). Due to uniformity of the STAR detector, $A_0$ is very close to unity, with only a few \% deviation depending on the centrality, $p_{\rm T}$, and $\eta$.
Owing to the overall system vorticity, the polarization becomes non-zero even for taking average over all hyperons in the phase-space, referred to as global polarization.

The polarization signal was extracted using the invariant mass method~\cite{STAR:2018gyt}.  
This requires measuring the mean value of $\sin(\Psi_{1}-\phi_{\rm B}^{\ast})$ as a function of the invariant mass $M_{\rm inv}$:
\begin{equation}
\begin{split}
\langle \sin(\Psi_{1}-\phi_{\rm B}^{*}) \rangle^{\rm obs}  =  (1-f^{\rm Bg}(M_{\rm inv})) \langle\sin(\Psi_{1}-\phi_{\rm B}^{\ast})\rangle^{\rm Sg} \\ 
~+~f^{\rm Bg}(M_{\rm inv})\langle\sin(\Psi_{1}-\phi_{\rm B}^{\ast})\rangle^{\rm Bg}.
\end{split}\label{eq:IMmethod}
\end{equation} 
This method needs a measurement of the background fraction $f^{\rm Bg}(M_{inv})$ in both the signal (``Sg") and background (``Bg") regions. 
The $f^{\rm Bg}(M_{\rm inv})$ was estimated by fitting off-peak region of the invariant mass distribution with the second-order polynomial function.
In principle, the protons and pions used to reconstruct $\Lambda$s in the background region should not originate from the $\Lambda$ and therefore should not be correlated with the $\Lambda$ spin. 
However, an improper pairing, e.g. of daughter protons matched with pions that did not come from the $\Lambda$ decay, could result in a non-zero background contribution. 
The signal for the $\Lambda$ polarization could leak into the off-mass-peak region via protons which are real $\Lambda$ daughters being paired with pions that are not from the decay.
The data were also fitted assuming the background signal being zero. 
The contribution from the background to the polarization is found to be small and is considered as a source of systematic uncertainty in this analysis (see section~\ref{syst}).

Although the invariant mass method was used as the default method in this analysis, the event plane method was also tested as a systematic check~\cite{STAR:2013ayu,Borghini:2004ra}. 
In the event plane method, the numbers of $\Lambda$ and $\bar{\Lambda}$ were counted in each bin of ($\Psi_{1}-\phi_{\rm B}^{\ast}$) after subtracting the combinatorial background. 
Then the yields of $\Lambda$ and $\bar{\Lambda}$ as a function of $\Psi_{1}-\phi_{\rm B}^{\ast}$ were fitted with a sine function to obtain $\langle\sin(\Psi_{1}-\phi_{\rm B}^{\ast})\rangle^{\rm Sg}$. 
Alternatively, the $\langle\sin(\Psi_{1}-\phi_{\rm B}^{\ast})\rangle^{\rm Sg}$ can be directly extracted by measuring an average of the distribution in the hyperon mass window and then correcting for the purity (profile method). 
The results from these three methods are found to be consistent within uncertainties and the difference is taken as a systematic uncertainty.

\label{Pyc2section}

The polarization component along the system angular momentum direction could depend on the azimuthal angle of the hyperon relative to the reaction plane. 
Considering the symmetry of the system geometry, the polarization can be written as a function of the hyperon's azimuthal angle relative to the reaction plane angle $\phi_{\rm H}-\Psi_{\rm RP}$ as follows~\cite{Niida:2024ntm}:
\begin{eqnarray}    
\!\!\!\!\!\! P_{\rm H}(\phi_{\rm H}-\Psi_{\rm RP})
    \!\!\!\! &=& \!\!\!\! P_{y,c0}+2\sum_{n=1}^{\infty} P_{y,c(2n)}\cos[2n(\phi_{\rm H}-\Psi_{\rm RP})] \\
    \label{equ:Pyc2}
    \!\!\!\! &\approx& \!\!\!\! P_{y,c0}  + 2P_{y,c2}\cos[2(\phi_{\rm H}-\Psi_{\rm RP})] \label{eq:Pyc0c2}.
\end{eqnarray}
Study of the coefficient $P_{y,2}$ is of particular interest as theoretical models predict differently the azimuthal modulation of $P_y$ (see more discussion in Sec.~\ref{sec:Pycn}). 

By considering the effect of detector acceptance as well as elliptic flow, the polarization coefficients $P_{y,c0}$ and $P_{y,c2}$, coupled with elliptic flow $v_2$ of hyperons, can be written as~\cite{Niida:2024ntm}:
\begin{eqnarray}
& &\hspace{-35pt} \langle\sin(\Psi_1-\phi_{\rm B}^{\ast})\rangle \nonumber \\
& &\hspace{-15pt} = \dfrac{\alpha_{\rm H}\pi}{8}\left[A_{0}\left(P_{y,c0}+2P_{y,c2}v_{2}\right)-A_{2}\left(P_{y,c2}+P_{y,c0}v_{2}\right)\right], \label{eq:Pyc0} \\
& &\hspace{-35pt} \langle\sin(\Psi_1-\phi_{\rm B}^{\ast})\cos[2(\phi_{\rm H}-\Psi_{RP})]\rangle \nonumber \\
& &\hspace{-15pt} = \dfrac{\alpha_{\rm H}\pi}{8}\left[A_{0}(P_{y,c2}+P_{y,c0}v_{2})-\dfrac{1}{2}A_{2}(P_{y,c0}+3P_{y,c2}v_{2})\right],\label{eq:Pyc2}
\end{eqnarray}
where $A_{0}$ and $A_{2}$ are acceptance correction factors defined as
\begin{eqnarray}
&&\hspace{-20pt} A_{0} = \dfrac{4}{\pi} \!\!\int\! \dfrac{d\Omega^{\ast}}{4\pi} \dfrac{d\phi_{\rm H}}{2\pi} A(\boldsymbol{p}_{\rm H}, \boldsymbol{p}^{\ast})\sin\theta_{\rm B}^{\ast}, \label{eq:A0}\\
&&\hspace{-20pt} A_{2} = \dfrac{4}{\pi} \!\!\int\! \dfrac{d\Omega^{\ast}}{4\pi} \dfrac{d\phi_{\rm H}}{2\pi} A(\boldsymbol{p}_{\rm H},\boldsymbol{p}^{\ast})\sin\theta_{\rm B}^{\ast}\cos[2(\phi_{\rm H}-\phi_{\rm B}^{\ast})].
\end{eqnarray}
The function $A(\boldsymbol{p}_{\rm H}, \boldsymbol{p}^{\ast})$ accounts for the detector acceptance and $\theta_{\rm B}^{\ast}$ is defined as the angle between the momentum of the decay baryon in the hyperon rest frame and the beam direction.
The factors $A_{0}$ and $A_{2}$ are extracted directly from the data. By solving Eqs.~\eqref{eq:Pyc0} and \eqref{eq:Pyc2} with the measurement of $v_2$, one can obtain the dependence of the polarization on the hyperon’s emission azimuthal angle with respect to the second harmonic event plane, the polarization coefficients $P_{y,c0}$ and $P_{y,c2}$ (method-1).

If the higher-order acceptance correction information on $\theta^{\ast}_{\rm B}$ is included in the measurement, one obtains another set of equations (method-2):
%
\begin{eqnarray}
& & \hspace{-35pt} \langle \sin(\Psi_1-\phi_{\rm B}^{\ast})\sin\theta_{\rm B}^{\ast}\rangle \nonumber \\
& & \hspace{-15pt} = \dfrac{\alpha_{\rm H}}{3}\left[\tilde{A}_{0}\left(P_{y,c0}+2P_{y,c2}v_{2}\right)-\tilde{A}_{2}\left(P_{y,c2}+P_{y,c0}v_{2}\right)\right], \label{eq:Pyc0tl}\\
& & \hspace{-35pt} \langle \sin(\Psi_1-\phi_{\rm B}^{\ast})\sin\theta_{\rm B}^{\ast}\cos[2(\phi_{\rm H}-\Psi_{RP})]\rangle \nonumber \\
& & \hspace{-15pt} = \dfrac{\alpha_{\rm H}}{3}\left[\tilde{A}_{0}(P_{y,c2}+P_{y,c0}v_{2})-\dfrac{1}{2}\tilde{A}_{2}(P_{y,c0}+3P_{y,c2}v_{2})\right] \label{eq:Pyc2tl},
\end{eqnarray}
with acceptance correction factors as follows:
\begin{eqnarray}
&&\hspace{-20pt} \tilde{A}_{0} \!=\! \dfrac{4}{\pi} \!\!\int\! \dfrac{d\Omega^{\ast}}{4\pi} \dfrac{d\phi_{\rm H}}{2\pi} A(\boldsymbol{p}_{\rm H}, \boldsymbol{p}^{\ast})\sin^2\theta_{\rm B}^{\ast}, \\
&&\hspace{-20pt} \tilde{A}_{2} \!=\! \dfrac{4}{\pi} \!\!\int\! \dfrac{d\Omega^{\ast}}{4\pi} \dfrac{d\phi_{\rm H}}{2\pi} A(\boldsymbol{p}_{\rm H},\boldsymbol{p}^{\ast})\sin^2\theta_{\rm B}^{\ast}\cos[2(\phi_{\rm H}-\phi_{\rm B}^{\ast})].
\end{eqnarray}
These two methods were tested to measure the polarization coefficient as presented in the following section.

In Eqs.~\eqref{eq:Pyc2} and \eqref{eq:Pyc2tl}, the second-order event plane determined by the TPC was used as a proxy of $\Psi_{\rm RP}$. Note that the tracks used for \lam (\alam) reconstruction were excluded from the event plane calculation to avoid a self-correlation.

\subsection{Systematic uncertainties}
\label{syst}

The systematic uncertainties were estimated by comparing the results obtained with different methods for the signal extraction and background contribution. 
Below, we describe each systematic source and provide typical values of relative uncertainty.

As mentioned in Sec.~\ref{Polarization_measurement}, three different techniques were used to extract the polarization signal. 
We use the results obtained with the invariant mass method as the default for $\Lambda$ global polarization. 
The difference in results from the event plane method was found to be 8\% and used in the evaluation  of the systematic uncertainty of the measurement. 
For the $\Xi^{-}$ global polarization, the event plane method was used as the default, and the difference in results from the profile method was found to be 15\%.

The uncertainty related to the background contribution was studied by changing the nonzero background assumption (used as the default) in Eq.~\ref{eq:IMmethod}. 
The difference of the zero-background assumption from the default in the invariant method was included in the systematic uncertainty ($\sim$7\% for \lam).
The difference in results under variation of the mass window for particles of interest from 3$\sigma$ to 2$\sigma$ in event plane method, was also included in the systematic uncertainty ($\sim$12\% for $\Xi$).
The uncertainty in the decay parameter $\alpha_{\rm H}$ was propagated through the analysis, resulting in a systematic uncertainty of $\sim$1.9\% for \lam and $\sim$2.6\% for $\Xi$, respectively~\cite{ParticleDataGroup:2022pth}.
The systematic uncertainties from the event plane resolution and acceptance effect are found to be small (below 0.1\%). 
The final systematic uncertainties were calculated by taking the square root of quadratic sum of the differences between the default condition and each systematic source. 

We note that the feed-down contributions reduce the polarization for primary \lam polarization by $\sim$10-15\% and enhance that for primary $\Xi$ by $\sim$25\%~\cite{Becattini:2016gvu,Karpenko:2016jyx,Xia:2019fjf,Becattini:2019ntv,Li:2021zwq}. 
As the correction for the feed-down contribution is model-dependent, we have chosen to present the polarization for the inclusive \lam and $\Xi$ samples in this study.

\section{Results and Discussions}

\subsection{$\Lambda$ global polarization}
\begin{figure}[H]
    \centering
    \includegraphics[width=0.9\linewidth]{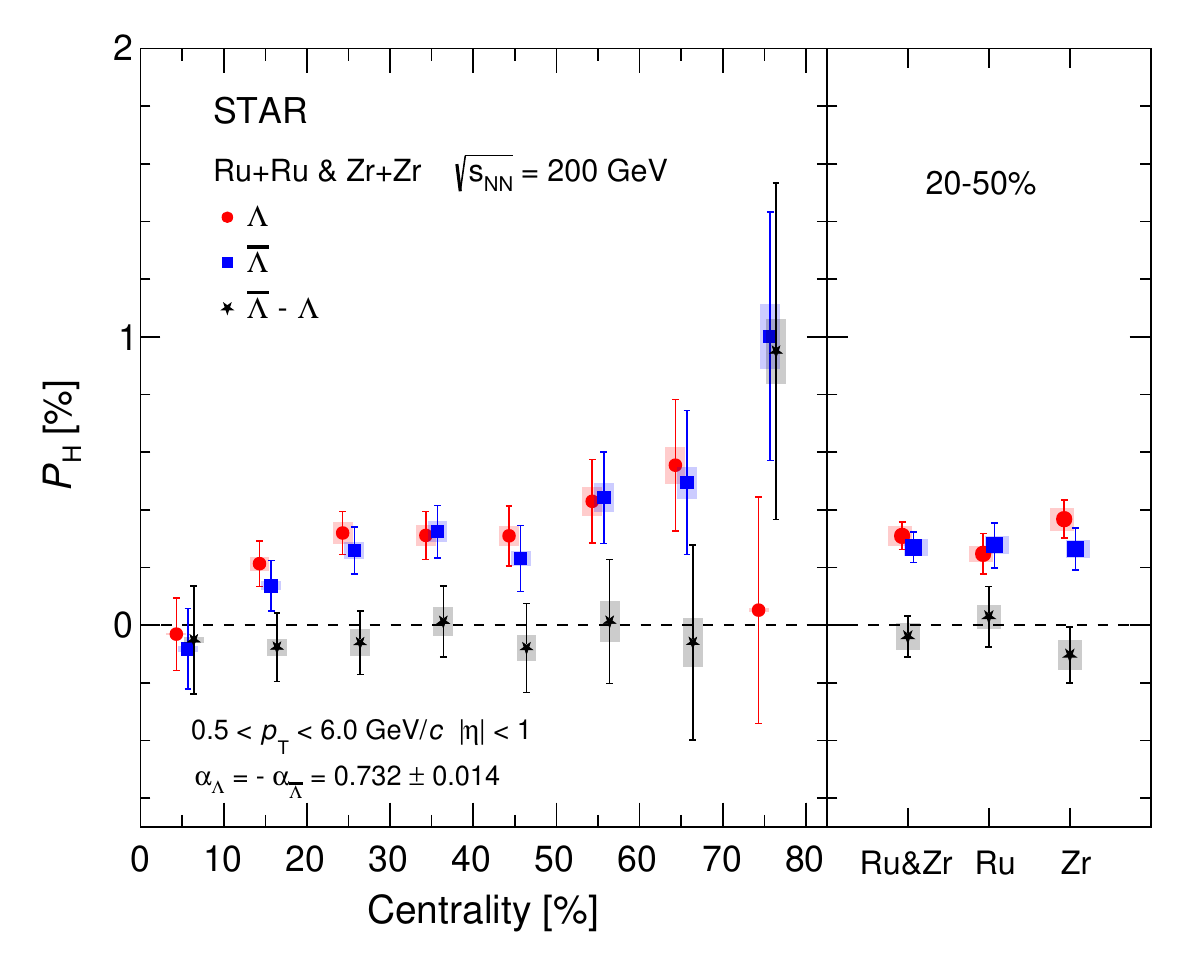}
    \vspace{-3mm}
    \caption{Global polarization of $\Lambda$ and $\bar\Lambda$ and their difference as a function of centrality in Ru+Ru and Zr+Zr collisions at \snn = 200 GeV in the left panel. 
    Right panel shows results for 20-50\% centrality window. Shaded boxes and vertical lines represent systematic and statistical uncertainties. 
    The data points are slightly shifted along the horizontal axis for visibility. }
    \label{fig:GlobalP_Lam_AntiLam}
\end{figure}

Figure~\ref{fig:GlobalP_Lam_AntiLam} presents the global polarization of $\Lambda$ and $\bar\Lambda$ and their difference as a function of centrality in Ru+Ru and Zr+Zr collisions at \snn = 200 GeV in the left panel, where the two system are combined.
The global polarization increases from central to peripheral collisions, following an increase of the thermal vorticity as expected. In the right panel, the results averaged over 20-50\% centrality interval are plotted: $P_{\Lambda} (\%) = 0.310\pm0.048({\rm stat.})\pm0.036({\rm syst.})$, $P_{\bar\Lambda} (\%) = 0.270\pm0.053({\rm stat.})\pm0.030({\rm syst.})$ for the two systems combined. The global polarization was also checked in Ru+Ru and Zr+Zr systems separately for 20-50\% centrality as shown in the right panel of Fig.~\ref{fig:GlobalP_Lam_AntiLam} and found be consistent with each other.

The difference in global polarization between \lam and \alam, $\Delta P_{\Lambda} = P_{\bar\Lambda}$ - $P_{\Lambda}$, can be used to study a possible contribution from the magnetic field. 
Assuming local thermal equilibrium, the late-stage magnetic field can be estimated as $B \approx T_f \Delta P_{\Lambda}/(2|\mu_{\Lambda}|)$~\cite{Becattini:2016gvu,Muller:2018ibh} where $T_f$ is the temperature when {\lam}s are emitted and $\mu_{\Lambda}$ is the magnetic moment of \lam. 
Our measurements show no significant difference between $\Lambda$ and $\bar\Lambda$ in the Ru+Ru and Zr+Zr collisions as well as the combined results. 
The baseline of the polarization difference without the magnetic field effect may not be zero, since other factors such as different space-time distributions, freeze-out conditions, the chemical potential, 
could contribute to the polarization difference in either positive or negative ways~\cite{STAR:2023nvo,Becattini:2016gvu,Fang:2016vpj,Csernai:2018yok,Vitiuk:2019rfv}.
One could expect larger $\Delta P_{\Lambda}$ in Ru+Ru than in Zr+Zr, if there is any contribution of the spin-magnetic coupling. 
The $\Delta P_{\Lambda}$ from each isobar collisions is found to be consistent with zero within the current precision, $\Delta P_{\Lambda}^{\rm Ru}-\Delta 
P_{\Lambda}^{\rm Zr}~(\%) = 0.133\pm0.143({\rm stat.})\pm0.066({\rm syst.})$.
Thus, in the following discussion, the two isobars were combined to increase the statistical precision.
\begin{figure}[htb]
    \centering
    \includegraphics[width=0.85\linewidth]{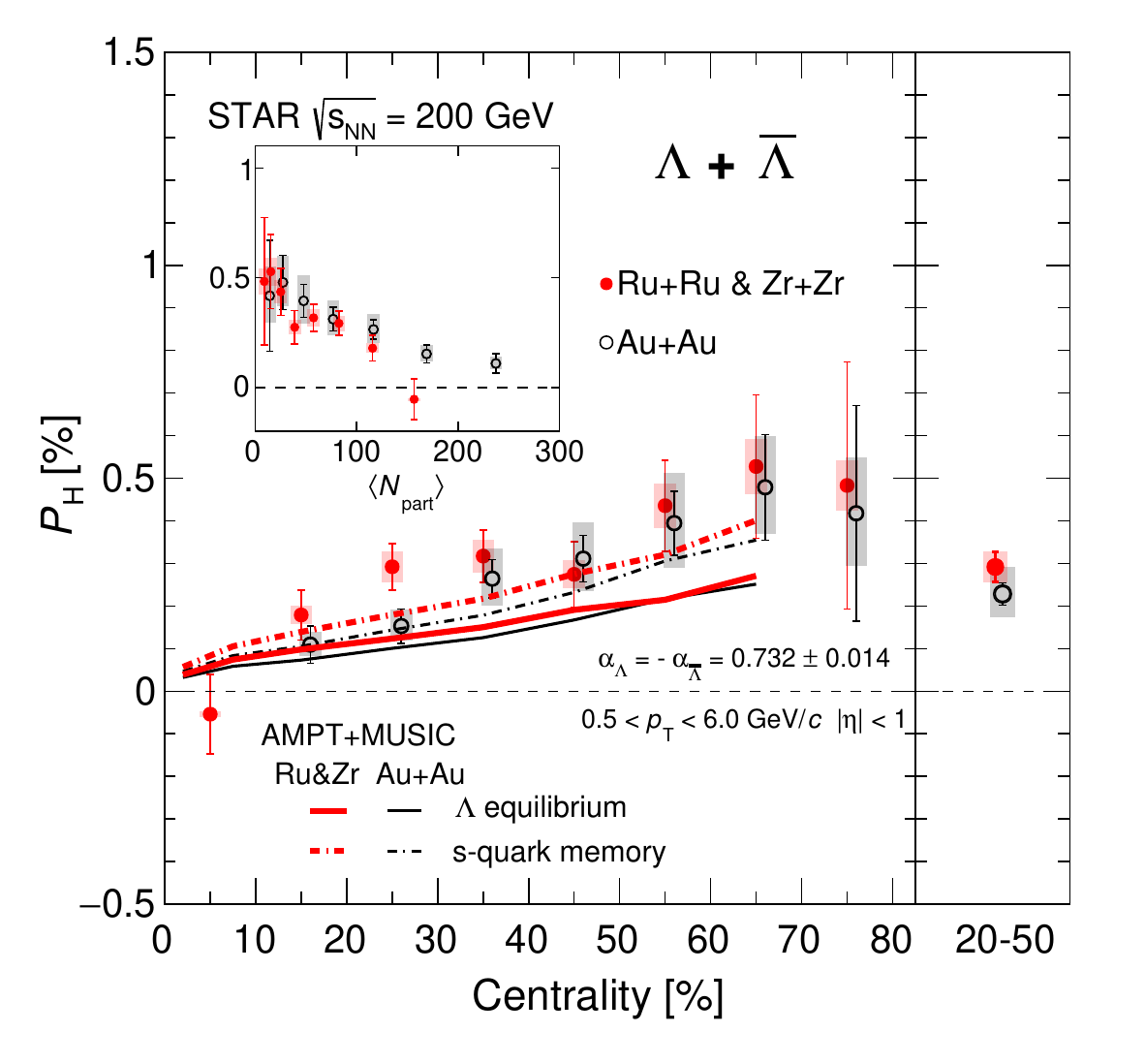}
    \vspace{-3mm}
    \caption{ Global polarization of $\Lambda$+$\bar\Lambda$ as a function of centrality in combined Ru+Ru and Zr+Zr collisions at \snn = 200 GeV (left panel) and results averaged over 20-50\% centrality (right panel). 
    Results for $\Lambda$+$\bar\Lambda$ measurements in Au+Au collisions at \snn = 200 GeV~\cite{STAR:2018gyt} are shown for comparison. 
    The inset presents the same data plotted as a function of the average number of participants $\langle N_{\rm part} \rangle$. 
    Solid and dashed-dotted lines show calculations from the hydrodynamic model (MUSIC) with the AMPT initial condition~\cite{Fu:2021pok} for the isobar collisions with two different scenarios (see texts). The data points are slightly shifted along the horizontal axis for visibility.}
    \label{fig:GlobalP_SystemSize}
\end{figure}

\begin{figure}[H]
    \centering
    \includegraphics[width=0.85\linewidth]{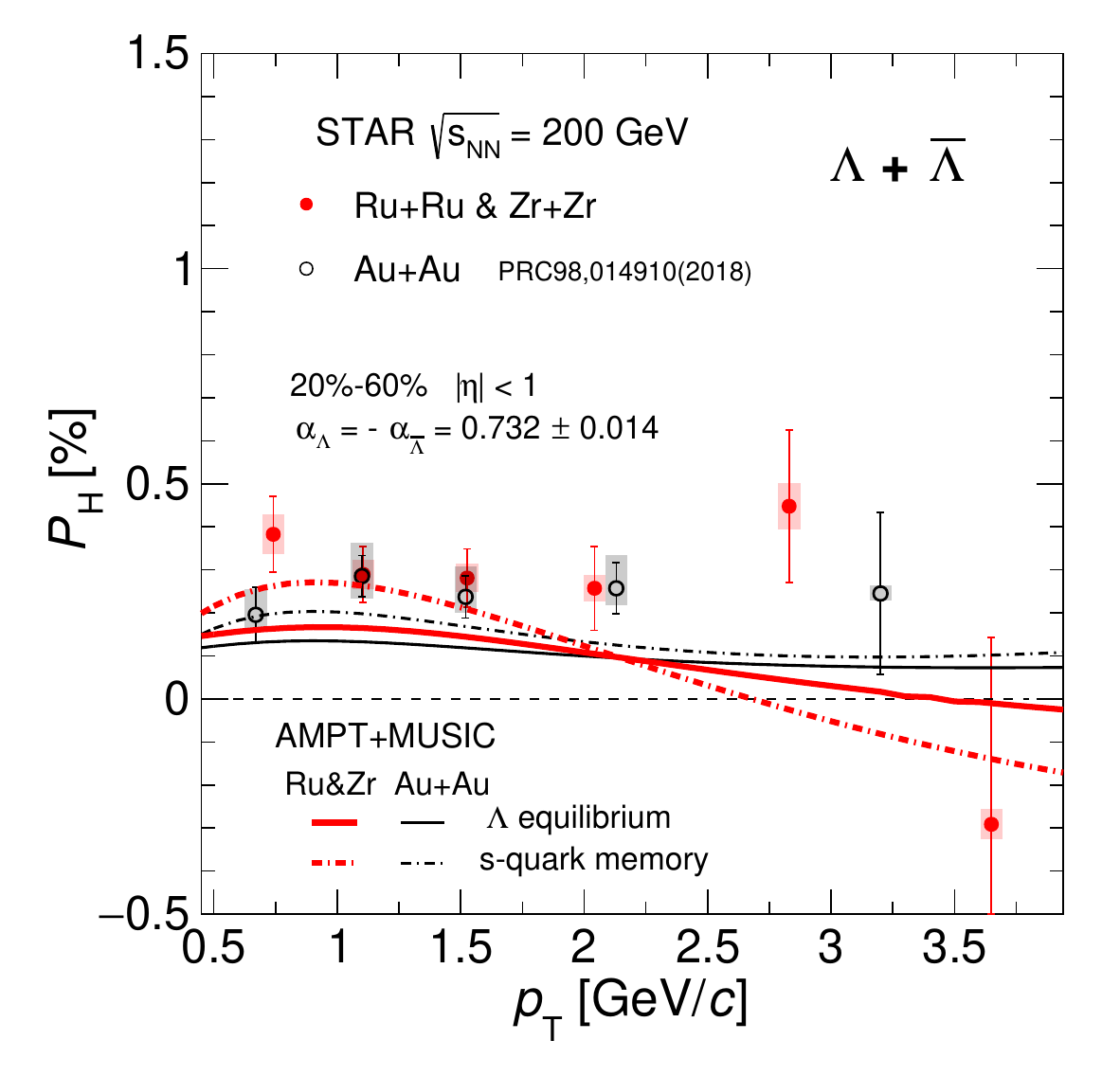}
    \vspace{-3mm}
    \caption{ Global polarization of $\Lambda$+$\bar\Lambda$ as a function of $p_{\rm T}$ for 20\%-60\% centrality in combined Ru+Ru and Zr+Zr collisions at \snn = 200 GeV, comparing with results for Au+Au collisions at \snn = 200 GeV~\cite{STAR:2018gyt}. 
    Shaded boxes and vertical lines show systematic and statistical uncertainties, respectively.  
    See Fig.~\ref{fig:GlobalP_SystemSize} caption for the lines.}
    \label{fig:GlobalP_PtinRuZr}
\end{figure}

\begin{figure}[H]
    \centering
    \includegraphics[width=0.85\linewidth]{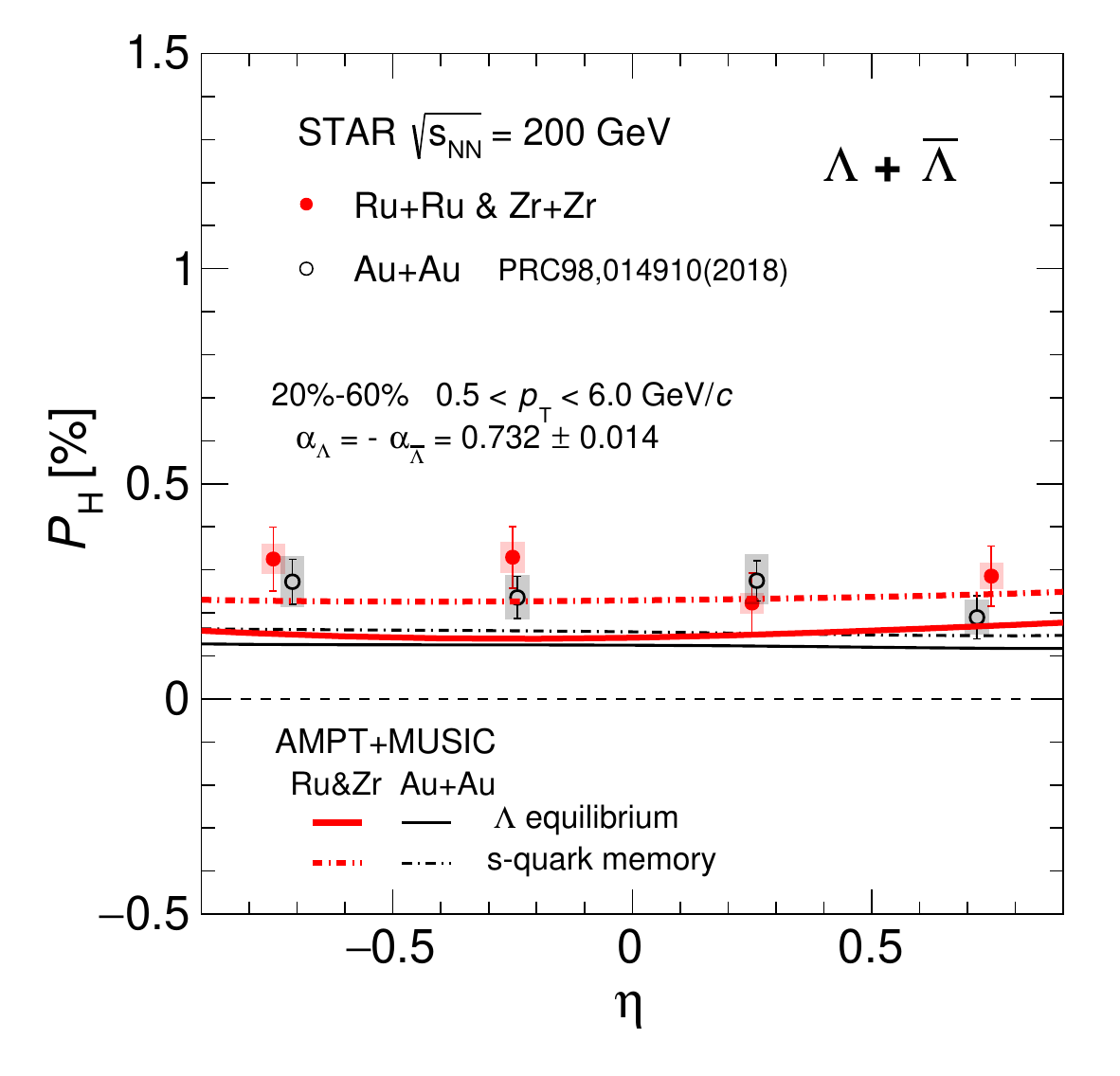}
    \vspace{-3mm}
    \caption{ Global polarization of $\Lambda$+$\bar\Lambda$ as a function of $\eta$ for 20\%-60\% centrality in combined Ru+Ru and Zr+Zr collisions at \snn = 200 GeV, comparing with results for Au+Au collisions at \snn = 200 GeV~\cite{STAR:2018gyt}. Shaded boxes and vertical lines show systematic and statistical uncertainties, respectively. 
    See Fig.~\ref{fig:GlobalP_SystemSize} caption for the lines.}
    \label{fig:GlobalP_EtainRuZr}
\end{figure}


Hydrodynamic calculations performed for different colliding systems reveal a larger global polarization in smaller systems at the same centrality~\cite{Shi:2017wpk,Alzhrani:2022dpi}.
This system-size dependence is attributed to a shorter fireball lifetime in smaller collision systems, where \lam hyperons are produced earlier when the vorticity is stronger before its dilution with time.
In Fig.~\ref{fig:GlobalP_SystemSize}, the comparison of $\Lambda$+$\bar\Lambda$ global polarization between isobar and Au+Au collisions at \snn = 200 GeV is shown.  
The results are comparable for the entire centrality range, with a hint for a slightly larger polarization in isobar collisions in mid centrality. 
Calculations of primary \lam polarization from a (3+1)D hydrodynamic model (MUSIC)~\cite{Fu:2021pok} with the initial condition from the AMPT model are compared to the data. 
Two scenarios are shown - ``\lam equilibrium" and ``s-quark memory", depending on the mass that is used to set the spin relaxation time along with the shear-induced polarization~\cite{Fu:2021pok,Becattini:2021iol}.
The former assumes that the spin of \lam immediately responds to the hydrodynamic gradients, while the latter considers that the \lam inherits the polarization from its constituent strange quark and is frozen since the hadronization.
The calculation with the s-quark memory describes the data better. 
The model shows a slight larger polarization in isobar collisions than in Au+Au collisions, and the difference  is comparable to the current statistical precision of the isobar results.
In the inset of Fig.~\ref{fig:GlobalP_SystemSize}, the data are plotted as a function of the average number of nucleon participants $\langle N_{\rm part}\rangle$ estimated from the Glauber model, where $N_{\rm part}$ averaged over the two systems were used.
The data seem to scale better with $N_{\rm part}$, possibly pointing to the significance of the system size in vorticity formation.

Figures~\ref{fig:GlobalP_PtinRuZr} and~\ref{fig:GlobalP_EtainRuZr} show $\Lambda+\bar\Lambda$ polarization dependence on $p_{\rm T}$ and $\eta$ for 20-60\% collision centrality. 
There is no obvious $p_{\rm T}$ or $\eta$ dependence, which is similar to the Au+Au results.
The calculation with the $s$-quark memory scenario shows a stronger $p_{\rm T}$ dependence in smaller collisions than the \lam equilibrium scenario.
The current precision of the measurement is not sufficient to constrain the models.
Although the theoretical models show substantial dependence on (pseudo)rapidity, especially in the far forward (backward) regions~\cite{Wu:2019eyi,Xie:2019jun,Liang:2019pst}, the data do not show any significant $\eta$ dependence within limited pseudorapidity region studied in this analysis.

\subsection{Azimuthal angle dependence}\label{sec:Pycn}

\begin{figure}[htb]
    \centering
    \includegraphics[width=0.85\linewidth]{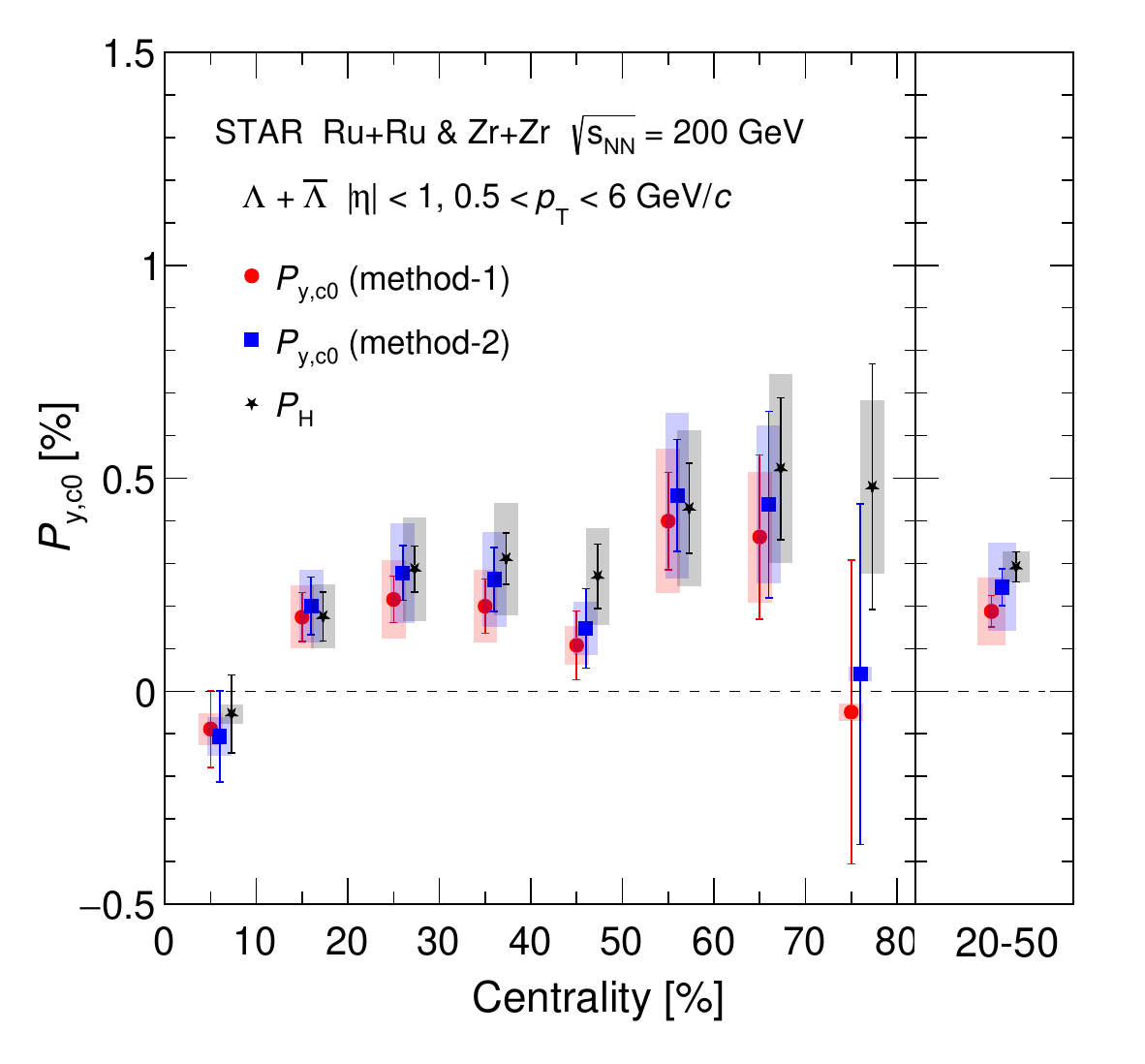}
    \vspace{-3mm}
    \caption{The polarization coefficients $P_{y,c0}$ of \lam+\alam obtained as a function of centrality (left panel) and for 20-50\% centrality (right panel) in isobar Ru+Ru and Zr+Zr collisions at \snn = 200 GeV, from the two methods (see texts). Azimuthally integrated results of the $P_{H}$ are shown for comparison. 
    The data points are slightly shifted along the horizontal axis for visibility.}
    \label{fig:Pyc0-cent}
\end{figure}
\vspace{3mm}
\vspace{-4mm}
\begin{figure}[thb]
    \centering
    \includegraphics[width=0.85\linewidth]{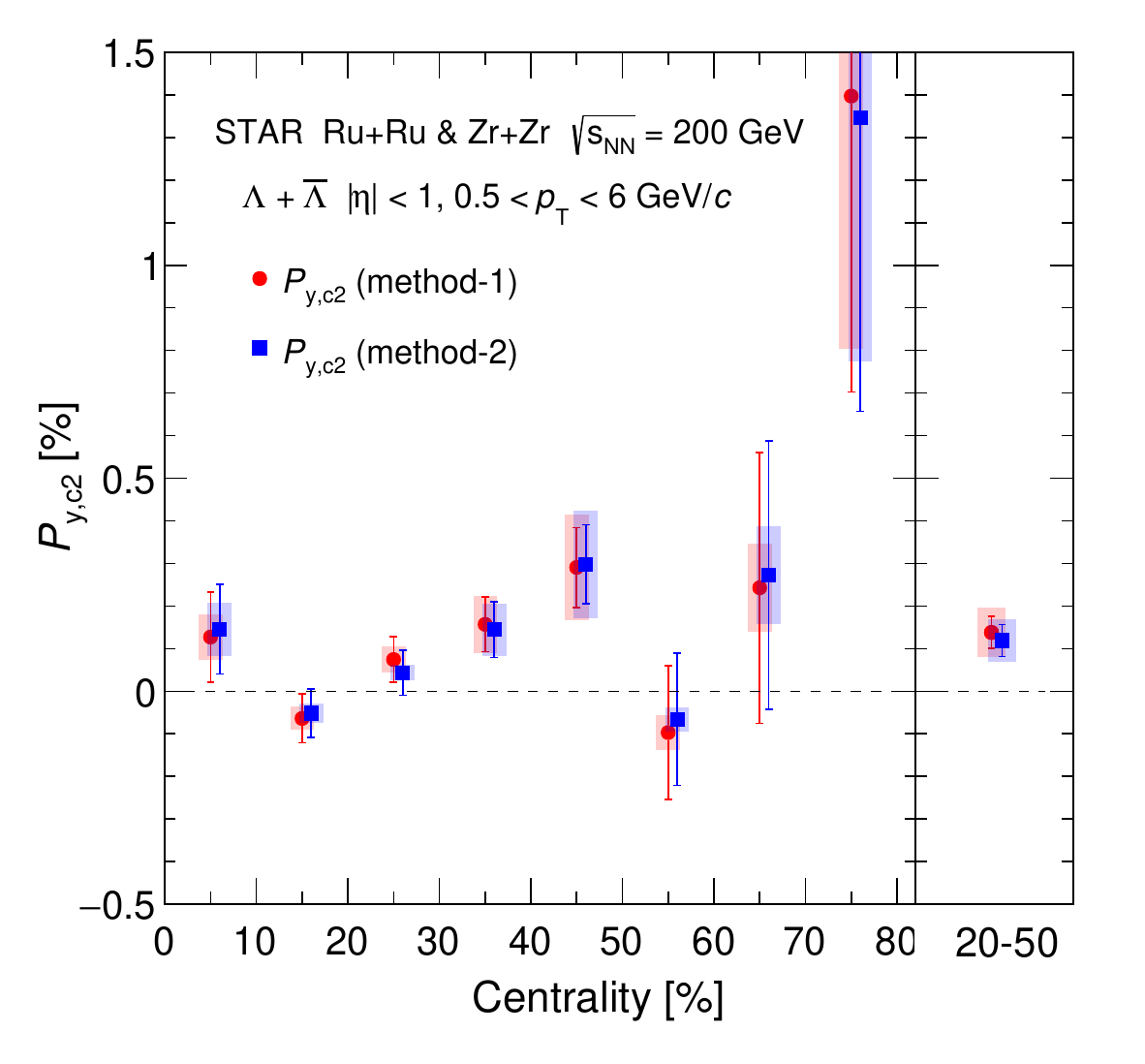}
    \vspace{-3mm}
    \caption{The polarization coefficients $P_{y,c2}$ of \lam+\alam obtained from the two methods as a function of centrality in isobar Ru+Ru and Zr+Zr collisions at \snn = 200 GeV. 
    The data points are slightly shifted along the horizontal axis for visibility.}
    \label{fig:Pyc2-cent}
\end{figure}

According to theoretical calculations in hydrodynamics and transport models, the polarization along the initial angular momentum direction of the system is expected to depend on azimuthal angle of hyperon emission. 
Calculations based on the thermal vorticity predict larger polarization in the out-of-plane direction~\cite{Becattini:2015ska,Russo:2016ueu,Wei:2018zfb,Fu:2020oxj}, while the sign of the modulation could be flipped, i.e. larger polarization along the in-plane direction, by incorporation of the thermal shear contribution. 
The latter depends on the detailed implementation of the calculations~\cite{Becattini:2021iol,Fu:2021pok,Yi:2021ryh,Alzhrani:2022dpi}. 
This situation is similar to a spin sign puzzle in measurements of polarization along the beam direction~\cite{STAR:2019erd,STAR:2023eck,ALICE:2021pzu}.

Figure~\ref{fig:Pyc0-cent} shows the coefficient $P_{y,c0}$ as defined in Eq.~\eqref{eq:Pyc0c2} and obtained with the two acceptance correction methods described in Eqs.~\eqref{eq:Pyc0} and \eqref{eq:Pyc0tl}. 
Both descriptions include the effect of the elliptic flow. 
In this study, the elliptic flow of \lam was measured with the same criteria as the polarization measurement, but it is confirmed that the results presented in another paper~\cite{strangev2} which applies different analysis criteria are consistent with the results used in this analysis.
%
%
The polarization results from the two methods are consistent with each other and agree with the $P_{\rm H}$ calculated using Eq.~\eqref{eq:PH}, where the polarization is extracted in the azimuthally-integrated fashion.

In Fig.~\ref{fig:Pyc2-cent}, the second-order Fourier cosine coefficient of the polarization $P_{y,c2}$ for \lam and \alam (see Eqs.~\eqref{eq:Pyc2} and \eqref{eq:Pyc2tl}) is presented as a function of centrality. 
The coefficient $P_{y.c2}$ is found to be positive at the 2.4$\sigma$ significance level, $P_{y.c2} (\%) = 0.138\pm0.038({\rm stat.})\pm0.058({\rm syst.})$, for mid-central collisions (20-50\%). 
This points to a larger polarization of hyperons emitted in-plane rather of those out-of-plane. 
Recent theoretical study shows the sensitivity of the azimuthal dependence of the polarization to the initial condition and bulk viscosity of the medium~\cite{Palermo:2024tza}. 
These results provide important new information for theoretical models aiming to model the polarization mechanism and to disentangle the contributions from thermal vorticity and shear-induced polarization.

\subsection{$\Xi^{-}+\bar\Xi^{+}$ global polarization}

The global polarization of charged $\Xi$ hyperons was studied by utilizing the two-step decay of $\Xi\rightarrow \Lambda\pi \rightarrow p\pi\pi$.
Two independent methods were used to measure $\Xi^{-}$($\bar\Xi^{+}$) polarization~\cite{STAR:2020xbm}. 
The first measures the $\Xi$ polarization directly by analyzing the daughter \lam distribution in the $\Xi$ rest frame.
The second method uses the granddaughter proton distribution in the daughter \lam rest frame to determine the \lam polarization and then uses polarization transfer coefficient ${\bm P}^\ast_{\Lambda}=C_{\Xi\Lambda}{\bm P}^\ast_{\Xi}$ to extract the parent $\Xi$ polarization. The polarization transfer coefficient from $\Xi$ to daughter $\Lambda$, $C_{\Xi\Lambda}$, is known to be $C_{\Xi\Lambda}=+0.944$~\cite{ParticleDataGroup:2022pth,STAR:2020xbm}.

\begin{figure}[htb]
    \centering
    \includegraphics[width=0.85\linewidth]{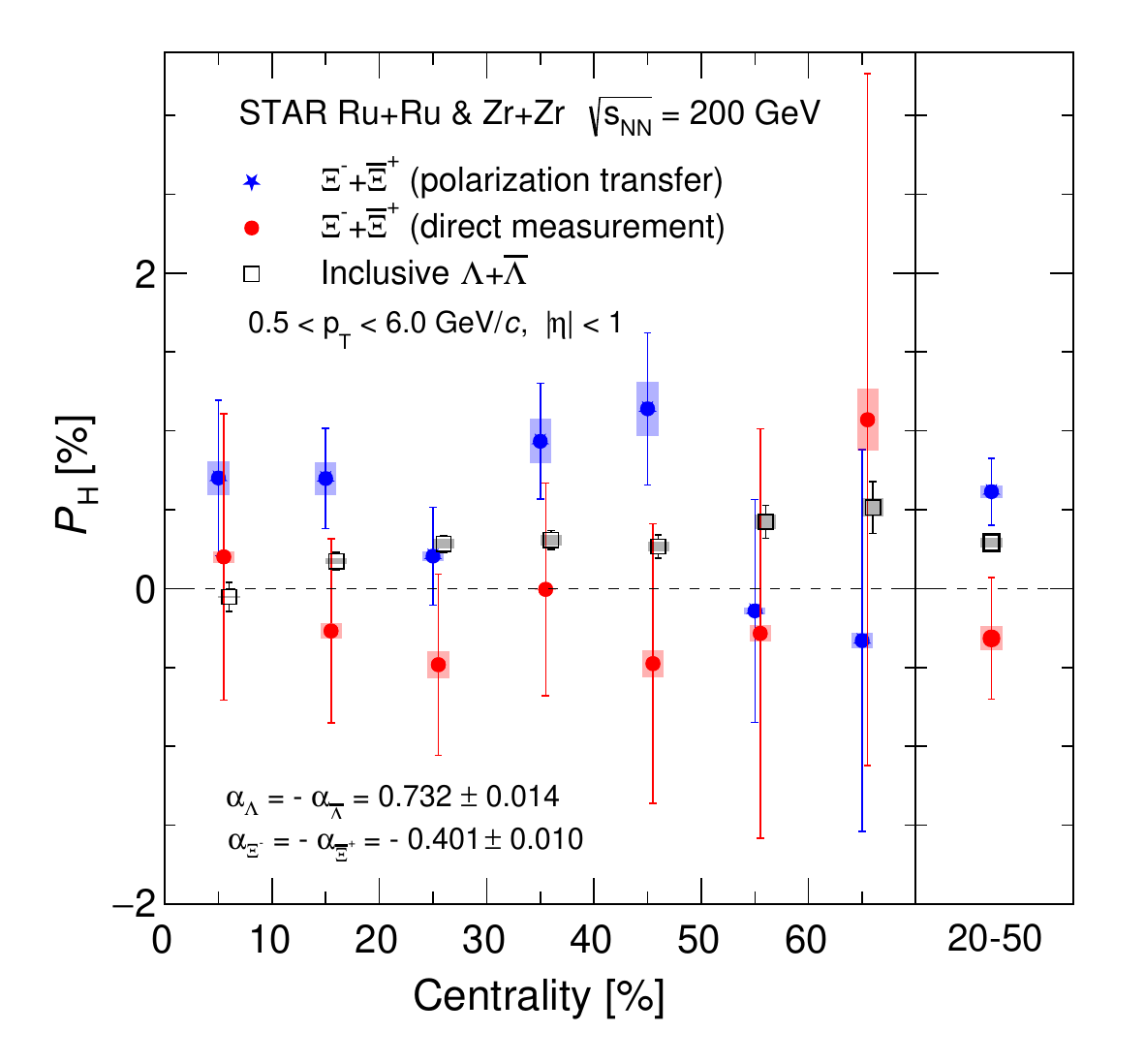}
    \vspace{-3mm}
    \caption{ Global polarization of $\Xi^{-}$+$\bar\Xi^{+}$ as a function of centrality in combined Ru+Ru and Zr+Zr collisions at \snn = 200 GeV. Results for the inclusive $\Lambda$+$\bar\Lambda$ measurements are shown for comparison. The data points are slightly shifted along the horizontal axis for visibility.}
    \label{fig:XiGlobalP}
\end{figure}

The combined $\Xi^{-}$ and $\bar\Xi^{+}$ global polarization results are shown in Fig.~\ref{fig:XiGlobalP}, with the inclusive \lam results for comparison.
The global polarization of $\Xi^{-}+\bar\Xi^{+}$ determined from the polarization transfer analysis is positive and found to be $P_{\Xi^{-}+\bar\Xi^{+}} (\%) = 0.61\pm0.21(\rm stat.)\pm0.04(\rm syst.)$ for 20-50\% centrality.
The $\Xi^{-}$+$\bar\Xi^{+}$ polarization determined from 
 the analysis of the daughter \lam distribution is found to be $P_{\Xi}(\%) = -0.32\pm0.39({\rm stat.})\pm0.08({\rm syst.})$ with large uncertainty in part due to a smaller value of $|\alpha_{\Xi}|$ compared to $|\alpha_{\Lambda}|$, which leads to smaller sensitivity of the measurement.
The results between the two methods are expected to be the same and the difference is $\sim2\sigma$ level for 20-50\% centrality. 
Considering the better sensitivity of the measurement with the polarization transfer, $P_\Xi$ results show a hint of larger polarization than the inclusive \lam.
The possible hierarchy of $P_{\Xi}>P_{\Lambda}$ is consistent with our previous measurement in Au+Au collisions~\cite{STAR:2020xbm} and could be explained by the thermal vorticity picture for spin-1/2 particles and feed-down effects~\cite{Li:2021zwq}.

\section{Summary}

We presented the results of global polarization measurements for $\Lambda$, $\bar\Lambda$, $\Xi^-$, and $\Xi^+$ hyperons in Ru+Ru and Zr+Zr collisions at \snn = 200 GeV. 
Global polarization of $\Lambda$+$\bar\Lambda$ shows no significant difference between isobar and Au+Au collisions, indicating no obvious collision system size dependence. 
Within the uncertainties, our results agree with calculations from hydrodynamic models incorporating the shear-induced polarization. 
No significant difference between $\Lambda$ and $\bar\Lambda$ global polarization in Ru+Ru and Zr+Zr collisions is observed, consistent with little contribution of the magnetic-field-driven effect on the observed hyperon polarization within present statistical precision. 
The polarization was also studied as a function of the collision centrality, the hyperon's transverse momentum, pseudorapidity, and azimuthal angle. 
The polarization was found to increase towards peripheral collisions, as expected from theoretical calculations, but shows no significant dependence on pseudorapidity or transverse momentum. 
The results on azimuthal angle dependence indicates larger polarization in the in-plane direction with 2.4$\sigma$ significance, which is an important input to understand the contribution of the shear-induced polarization.
Furthermore, $\Xi^-$ and $\Xi^+$ hyperon global polarization via the measurement of daughter \lam polarization indicates positive polarization (2.9$\sigma$ significance in 20-50\% centrality) as expected from the global vorticity picture. 

\section*{Acknowledgements}
\input{acknowledgment}



\bibliographystyle{elsarticle-num}
\bibliography{refs}

\end{document}

%% file: macro.tex
\usepackage{graphicx}   
\usepackage{float} 
\usepackage{xspace}     
\usepackage{url}
\usepackage[mathlines]{lineno}
\usepackage{bm}
\usepackage{amsmath}
\usepackage{xcolor}
\usepackage[driverfallback=dvipdfm]{hyperref}
\usepackage{ulem}
\usepackage{subfigure}
\hypersetup{
	colorlinks = true,
	citecolor= blue,
	linkbordercolor = cyan
}

\newcommand{\lam}{\mbox{$\Lambda$}\xspace}
\newcommand{\alam}{\mbox{$\bar{\Lambda}$}\xspace}

\def \be {\begin{equation}}
\def \ee {\end{equation}} 
\def \bea {\begin{eqnarray}}
\def \eea {\end{eqnarray}}

\def \snn {$\sqrt{s_{_{\rm NN}}}$\xspace} 
\usepackage{color}
\definecolor{orange}{cmyk}{0.,0.353,1.,0.}    

\usepackage{lineno}
\newcommand*\patchAmsMathEnvironmentForLineno[1]{
  \expandafter\let\csname old#1\expandafter\endcsname\csname #1\endcsname
  \expandafter\let\csname oldend#1\expandafter\endcsname\csname end#1\endcsname
  \renewenvironment{#1}
     {\linenomath\csname old#1\endcsname}
     {\csname oldend#1\endcsname\endlinenomath}}
\newcommand*\patchBothAmsMathEnvironmentsForLineno[1]{
  \patchAmsMathEnvironmentForLineno{#1}
  \patchAmsMathEnvironmentForLineno{#1*}}
\AtBeginDocument{
\patchBothAmsMathEnvironmentsForLineno{equation}
\patchBothAmsMathEnvironmentsForLineno{align}
\patchBothAmsMathEnvironmentsForLineno{flalign}
\patchBothAmsMathEnvironmentsForLineno{alignat}
\patchBothAmsMathEnvironmentsForLineno{gather}
\patchBothAmsMathEnvironmentsForLineno{multline}
}

%% file: acknowledgment.tex

We thank the RHIC Operations Group and SDCC at BNL, the NERSC Center at LBNL, and the Open Science Grid consortium for providing resources and support.  This work was supported in part by the Office of Nuclear Physics within the U.S. DOE Office of Science, the U.S. National Science Foundation, National Natural Science Foundation of China, Chinese Academy of Science, the Ministry of Science and Technology of China and the Chinese Ministry of Education, NSTC Taipei, the National Research Foundation of Korea, Czech Science Foundation and Ministry of Education, Youth and Sports of the Czech Republic, Hungarian National Research, Development and Innovation Office, New National Excellency Programme of the Hungarian Ministry of Human Capacities, Department of Atomic Energy and Department of Science and Technology of the Government of India, the National Science Centre and WUT ID-UB of Poland, the Ministry of Science, Education and Sports of the Republic of Croatia, German Bundesministerium f\"ur Bildung, Wissenschaft, Forschung and Technologie (BMBF), Helmholtz Association, Ministry of Education, Culture, Sports, Science, and Technology (MEXT), Japan Society for the Promotion of Science (JSPS) and Agencia Nacional de Investigaci\'on y Desarrollo (ANID) of Chile.

%% file: main-PLB.bbl
\begin{thebibliography}{10}
\expandafter\ifx\csname url\endcsname\relax
  \def\url#1{\texttt{#1}}\fi
\expandafter\ifx\csname urlprefix\endcsname\relax\def\urlprefix{URL }\fi
\expandafter\ifx\csname href\endcsname\relax
  \def\href#1#2{#2} \def\path#1{#1}\fi

\bibitem{1}
Z.-T. Liang, X.-N. Wang, {Globally polarized quark-gluon plasma in non-central A+A collisions}, Phys. Rev. Lett. 94 (2005) 102301, [Erratum: Phys.Rev.Lett. 96, 039901 (2006)].
\newblock \href {http://arxiv.org/abs/nucl-th/0410079} {\path{arXiv:nucl-th/0410079}}, \href {https://doi.org/10.1103/PhysRevLett.94.102301} {\path{doi:10.1103/PhysRevLett.94.102301}}.

\bibitem{Voloshin:2004ha}
S.~A. Voloshin, {Polarized secondary particles in unpolarized high energy hadron-hadron collisions? }\href {http://arxiv.org/abs/nucl-th/0410089} {\path{arXiv:nucl-th/0410089}}.

\bibitem{Liang:2004xn}
Z.-T. Liang, X.-N. Wang, {Spin alignment of vector mesons in non-central A+A collisions}, Phys. Lett. B 629 (2005) 20--26.
\newblock \href {http://arxiv.org/abs/nucl-th/0411101} {\path{arXiv:nucl-th/0411101}}, \href {https://doi.org/10.1016/j.physletb.2005.09.060} {\path{doi:10.1016/j.physletb.2005.09.060}}.

\bibitem{2}
F.~Becattini, F.~Piccinini, J.~Rizzo, {Angular momentum conservation in heavy ion collisions at very high energy}, Phys. Rev. C 77 (2008) 024906.
\newblock \href {http://arxiv.org/abs/0711.1253} {\path{arXiv:0711.1253}}, \href {https://doi.org/10.1103/PhysRevC.77.024906} {\path{doi:10.1103/PhysRevC.77.024906}}.

\bibitem{Gao:2007bc}
J.-H. Gao, S.-W. Chen, W.-T. Deng, Z.-T. Liang, Q.~Wang, X.-N. Wang, {Global quark polarization in non-central A+A collisions}, Phys. Rev. C 77 (2008) 044902.
\newblock \href {http://arxiv.org/abs/0710.2943} {\path{arXiv:0710.2943}}, \href {https://doi.org/10.1103/PhysRevC.77.044902} {\path{doi:10.1103/PhysRevC.77.044902}}.

\bibitem{STAR:2017ckg}
L.~Adamczyk, et~al., {Global $\Lambda$ hyperon polarization in nuclear collisions: evidence for the most vortical fluid}, Nature 548 (2017) 62--65.
\newblock \href {http://arxiv.org/abs/1701.06657} {\path{arXiv:1701.06657}}, \href {https://doi.org/10.1038/nature23004} {\path{doi:10.1038/nature23004}}.

\bibitem{STAR:2022fan}
M.~S. Abdallah, et~al., {Pattern of global spin alignment of \ensuremath{\phi} and K$^{*0}$ mesons in heavy-ion collisions}, Nature 614~(7947) (2023) 244--248.
\newblock \href {http://arxiv.org/abs/2204.02302} {\path{arXiv:2204.02302}}, \href {https://doi.org/10.1038/s41586-022-05557-5} {\path{doi:10.1038/s41586-022-05557-5}}.

\bibitem{STAR:2020xbm}
J.~Adam, et~al., {Global Polarization of $\Xi$ and $\Omega$ Hyperons in Au+Au Collisions at $\sqrt {s_{NN}}$ = 200 GeV}, Phys. Rev. Lett. 126~(16) (2021) 162301.
\newblock \href {http://arxiv.org/abs/2012.13601} {\path{arXiv:2012.13601}}, \href {https://doi.org/10.1103/PhysRevLett.126.162301} {\path{doi:10.1103/PhysRevLett.126.162301}}.

\bibitem{STAR:2007ccu}
B.~I. Abelev, et~al., {Global polarization measurement in Au+Au collisions}, Phys. Rev. C 76 (2007) 024915, [Erratum: Phys.Rev.C 95, 039906 (2017)].
\newblock \href {http://arxiv.org/abs/0705.1691} {\path{arXiv:0705.1691}}, \href {https://doi.org/10.1103/PhysRevC.76.024915} {\path{doi:10.1103/PhysRevC.76.024915}}.

\bibitem{STAR:2018gyt}
J.~Adam, et~al., {Global polarization of $\Lambda$ hyperons in Au+Au collisions at $\sqrt{s_{_{NN}}}$ = 200 GeV}, Phys. Rev. C 98 (2018) 014910.
\newblock \href {http://arxiv.org/abs/1805.04400} {\path{arXiv:1805.04400}}, \href {https://doi.org/10.1103/PhysRevC.98.014910} {\path{doi:10.1103/PhysRevC.98.014910}}.

\bibitem{ALICE:2019onw}
S.~Acharya, et~al., {Global polarization of $\Lambda \bar \Lambda$ hyperons in Pb-Pb collisions at $\sqrt {s_{NN}}$ = 2.76 and 5.02 TeV}, Phys. Rev. C 101~(4) (2020) 044611, [Erratum: Phys.Rev.C 105, 029902 (2022)].
\newblock \href {http://arxiv.org/abs/1909.01281} {\path{arXiv:1909.01281}}, \href {https://doi.org/10.1103/PhysRevC.101.044611} {\path{doi:10.1103/PhysRevC.101.044611}}.

\bibitem{STAR:2021beb}
M.~S. Abdallah, et~al., {Global $\Lambda$-hyperon polarization in Au+Au collisions at $\sqrt {s_{NN}}$=3~GeV}, Phys. Rev. C 104~(6) (2021) L061901.
\newblock \href {http://arxiv.org/abs/2108.00044} {\path{arXiv:2108.00044}}, \href {https://doi.org/10.1103/PhysRevC.104.L061901} {\path{doi:10.1103/PhysRevC.104.L061901}}.

\bibitem{STAR:2023nvo}
M.~I. Abdulhamid, et~al., {Global polarization of \ensuremath{\Lambda} and \ensuremath{\Lambda}\textasciimacron{} hyperons in Au+Au collisions at sNN=19.6 and 27 GeV}, Phys. Rev. C 108~(1) (2023) 014910.
\newblock \href {http://arxiv.org/abs/2305.08705} {\path{arXiv:2305.08705}}, \href {https://doi.org/10.1103/PhysRevC.108.014910} {\path{doi:10.1103/PhysRevC.108.014910}}.

\bibitem{HADES:2022enx}
R.~Abou~Yassine, et~al., {Measurement of global polarization of \ensuremath{\Lambda} hyperons in few-GeV heavy-ion collisions}, Phys. Lett. B 835 (2022) 137506.
\newblock \href {http://arxiv.org/abs/2207.05160} {\path{arXiv:2207.05160}}, \href {https://doi.org/10.1016/j.physletb.2022.137506} {\path{doi:10.1016/j.physletb.2022.137506}}.

\bibitem{Deng:2016gyh}
W.-T. Deng, X.-G. Huang, {Vorticity in Heavy-Ion Collisions}, Phys. Rev. C 93~(6) (2016) 064907.
\newblock \href {http://arxiv.org/abs/1603.06117} {\path{arXiv:1603.06117}}, \href {https://doi.org/10.1103/PhysRevC.93.064907} {\path{doi:10.1103/PhysRevC.93.064907}}.

\bibitem{Wei:2018zfb}
D.-X. Wei, W.-T. Deng, X.-G. Huang, {Thermal vorticity and spin polarization in heavy-ion collisions}, Phys. Rev. C 99~(1) (2019) 014905.
\newblock \href {http://arxiv.org/abs/1810.00151} {\path{arXiv:1810.00151}}, \href {https://doi.org/10.1103/PhysRevC.99.014905} {\path{doi:10.1103/PhysRevC.99.014905}}.

\bibitem{Guo:2021udq}
Y.~Guo, J.~Liao, E.~Wang, H.~Xing, H.~Zhang, {Hyperon polarization from the vortical fluid in low-energy nuclear collisions}, Phys. Rev. C 104~(4) (2021) L041902.
\newblock \href {http://arxiv.org/abs/2105.13481} {\path{arXiv:2105.13481}}, \href {https://doi.org/10.1103/PhysRevC.104.L041902} {\path{doi:10.1103/PhysRevC.104.L041902}}.

\bibitem{Bozek:2010bi}
P.~Bozek, I.~Wyskiel, {Directed flow in ultrarelativistic heavy-ion collisions}, Phys. Rev. C 81 (2010) 054902.
\newblock \href {http://arxiv.org/abs/1002.4999} {\path{arXiv:1002.4999}}, \href {https://doi.org/10.1103/PhysRevC.81.054902} {\path{doi:10.1103/PhysRevC.81.054902}}.

\bibitem{Ivanov:2018eej}
Y.~B. Ivanov, A.~A. Soldatov, {Vortex rings in fragmentation regions in heavy-ion collisions at $\sqrt{s_{NN}}=$ 39 GeV}, Phys. Rev. C 97~(4) (2018) 044915.
\newblock \href {http://arxiv.org/abs/1803.01525} {\path{arXiv:1803.01525}}, \href {https://doi.org/10.1103/PhysRevC.97.044915} {\path{doi:10.1103/PhysRevC.97.044915}}.

\bibitem{Wu:2019eyi}
H.-Z. Wu, L.-G. Pang, X.-G. Huang, Q.~Wang, {Local spin polarization in high energy heavy ion collisions}, Phys. Rev. Research. 1 (2019) 033058.
\newblock \href {http://arxiv.org/abs/1906.09385} {\path{arXiv:1906.09385}}, \href {https://doi.org/10.1103/PhysRevResearch.1.033058} {\path{doi:10.1103/PhysRevResearch.1.033058}}.

\bibitem{Ivanov:2019ern}
Y.~B. Ivanov, V.~D. Toneev, A.~A. Soldatov, {Estimates of hyperon polarization in heavy-ion collisions at collision energies $\sqrt{s_{NN}}=$ 4--40 GeV}, Phys. Rev. C 100~(1) (2019) 014908.
\newblock \href {http://arxiv.org/abs/1903.05455} {\path{arXiv:1903.05455}}, \href {https://doi.org/10.1103/PhysRevC.100.014908} {\path{doi:10.1103/PhysRevC.100.014908}}.

\bibitem{Ivanov:2020wak}
Y.~B. Ivanov, A.~A. Soldatov, {Correlation between global polarization, angular momentum, and flow in heavy-ion collisions}, Phys. Rev. C 102~(2) (2020) 024916.
\newblock \href {http://arxiv.org/abs/2004.05166} {\path{arXiv:2004.05166}}, \href {https://doi.org/10.1103/PhysRevC.102.024916} {\path{doi:10.1103/PhysRevC.102.024916}}.

\bibitem{Liang:2019pst}
Z.-T. Liang, J.~Song, I.~Upsal, Q.~Wang, Z.-B. Xu, {Rapidity dependence of global polarization in heavy ion collisions}, Chin. Phys. C 45~(1) (2021) 014102.
\newblock \href {http://arxiv.org/abs/1912.10223} {\path{arXiv:1912.10223}}, \href {https://doi.org/10.1088/1674-1137/abc065} {\path{doi:10.1088/1674-1137/abc065}}.

\bibitem{Jiang:2016woz}
Y.~Jiang, Z.-W. Lin, J.~Liao, {Rotating quark-gluon plasma in relativistic heavy ion collisions}, Phys. Rev. C 94~(4) (2016) 044910, [Erratum: Phys.Rev.C 95, 049904 (2017)].
\newblock \href {http://arxiv.org/abs/1602.06580} {\path{arXiv:1602.06580}}, \href {https://doi.org/10.1103/PhysRevC.94.044910} {\path{doi:10.1103/PhysRevC.94.044910}}.

\bibitem{McLerran:2013hla}
L.~McLerran, V.~Skokov, {Comments About the Electromagnetic Field in Heavy-Ion Collisions}, Nucl. Phys. A 929 (2014) 184--190.
\newblock \href {http://arxiv.org/abs/1305.0774} {\path{arXiv:1305.0774}}, \href {https://doi.org/10.1016/j.nuclphysa.2014.05.008} {\path{doi:10.1016/j.nuclphysa.2014.05.008}}.

\bibitem{Becattini:2016gvu}
F.~Becattini, I.~Karpenko, M.~Lisa, I.~Upsal, S.~Voloshin, {Global hyperon polarization at local thermodynamic equilibrium with vorticity, magnetic field and feed-down}, Phys. Rev. C 95~(5) (2017) 054902.
\newblock \href {http://arxiv.org/abs/1610.02506} {\path{arXiv:1610.02506}}, \href {https://doi.org/10.1103/PhysRevC.95.054902} {\path{doi:10.1103/PhysRevC.95.054902}}.

\bibitem{Muller:2018ibh}
B.~M\"uller, A.~Sch\"afer, {Chiral magnetic effect and an experimental bound on the late time magnetic field strength}, Phys. Rev. D 98~(7) (2018) 071902.
\newblock \href {http://arxiv.org/abs/1806.10907} {\path{arXiv:1806.10907}}, \href {https://doi.org/10.1103/PhysRevD.98.071902} {\path{doi:10.1103/PhysRevD.98.071902}}.

\bibitem{Xie:2019jun}
Y.~Xie, D.~Wang, L.~P. Csernai, {Fluid dynamics study of the $\varLambda $ polarization for Au + Au collisions at $\sqrt{s_{NN}}=200$ GeV}, Eur. Phys. J. C 80~(1) (2020) 39.
\newblock \href {http://arxiv.org/abs/1907.00773} {\path{arXiv:1907.00773}}, \href {https://doi.org/10.1140/epjc/s10052-019-7576-8} {\path{doi:10.1140/epjc/s10052-019-7576-8}}.

\bibitem{Shi:2017wpk}
S.~Shi, K.~Li, J.~Liao, {Searching for the Subatomic Swirls in the CuCu and CuAu Collisions}, Phys. Lett. B 788 (2019) 409--413.
\newblock \href {http://arxiv.org/abs/1712.00878} {\path{arXiv:1712.00878}}, \href {https://doi.org/10.1016/j.physletb.2018.09.066} {\path{doi:10.1016/j.physletb.2018.09.066}}.

\bibitem{Alzhrani:2022dpi}
S.~Alzhrani, S.~Ryu, C.~Shen, {\ensuremath{\Lambda} spin polarization in event-by-event relativistic heavy-ion collisions}, Phys. Rev. C 106~(1) (2022) 014905.
\newblock \href {http://arxiv.org/abs/2203.15718} {\path{arXiv:2203.15718}}, \href {https://doi.org/10.1103/PhysRevC.106.014905} {\path{doi:10.1103/PhysRevC.106.014905}}.

\bibitem{STAR:2019erd}
J.~Adam, et~al., {Polarization of $\Lambda$ ($\bar{\Lambda}$) hyperons along the beam direction in Au+Au collisions at $\sqrt{s_{_{NN}}}$ = 200 GeV}, Phys. Rev. Lett. 123~(13) (2019) 132301.
\newblock \href {http://arxiv.org/abs/1905.11917} {\path{arXiv:1905.11917}}, \href {https://doi.org/10.1103/PhysRevLett.123.132301} {\path{doi:10.1103/PhysRevLett.123.132301}}.

\bibitem{ALICE:2021pzu}
S.~Acharya, et~al., {Polarization of $\Lambda$ and $\bar \Lambda$ Hyperons along the Beam Direction in Pb-Pb Collisions at $\sqrt {s_{NN}}$=5.02\,\,TeV}, Phys. Rev. Lett. 128~(17) (2022) 172005.
\newblock \href {http://arxiv.org/abs/2107.11183} {\path{arXiv:2107.11183}}, \href {https://doi.org/10.1103/PhysRevLett.128.172005} {\path{doi:10.1103/PhysRevLett.128.172005}}.

\bibitem{STAR:2023eck}
M.~Abdulhamid, et~al., {Hyperon Polarization along the Beam Direction Relative to the Second and Third Harmonic Event Planes in Isobar Collisions at sNN=200\,\,GeV}, Phys. Rev. Lett. 131~(20) (2023) 202301.
\newblock \href {http://arxiv.org/abs/2303.09074} {\path{arXiv:2303.09074}}, \href {https://doi.org/10.1103/PhysRevLett.131.202301} {\path{doi:10.1103/PhysRevLett.131.202301}}.

\bibitem{Becattini:2020ngo}
F.~Becattini, M.~A. Lisa, {Polarization and Vorticity in the Quark\textendash{}Gluon Plasma}, Ann. Rev. Nucl. Part. Sci. 70 (2020) 395--423.
\newblock \href {http://arxiv.org/abs/2003.03640} {\path{arXiv:2003.03640}}, \href {https://doi.org/10.1146/annurev-nucl-021920-095245} {\path{doi:10.1146/annurev-nucl-021920-095245}}.

\bibitem{Becattini:2024uha}
F.~Becattini, M.~Buzzegoli, T.~Niida, S.~Pu, A.-H. Tang, Q.~Wang, {Spin polarization in relativistic heavy-ion collisions}, Int. J. Mod. Phys. E 33~(06) (2024) 2430006.
\newblock \href {http://arxiv.org/abs/2402.04540} {\path{arXiv:2402.04540}}, \href {https://doi.org/10.1142/S0218301324300066} {\path{doi:10.1142/S0218301324300066}}.

\bibitem{Niida:2024ntm}
T.~Niida, S.~A. Voloshin, {Polarization phenomenon in heavy-ion collisions}, Int. J. Mod. Phys. E 33~(09) (2024) 2430010.
\newblock \href {http://arxiv.org/abs/2404.11042} {\path{arXiv:2404.11042}}, \href {https://doi.org/10.1142/S0218301324300108} {\path{doi:10.1142/S0218301324300108}}.

\bibitem{Chen:2024aom}
J.~Chen, et~al., {Properties of the QCD matter: review of selected results from the relativistic heavy ion collider beam energy scan (RHIC BES) program}, Nucl. Sci. Tech. 35~(12) (2024) 214.
\newblock \href {http://arxiv.org/abs/2407.02935} {\path{arXiv:2407.02935}}, \href {https://doi.org/10.1007/s41365-024-01591-2} {\path{doi:10.1007/s41365-024-01591-2}}.

\bibitem{STAR:2021mii}
M.~Abdallah, et~al., {Search for the chiral magnetic effect with isobar collisions at $\sqrt {s_{NN}}$=200 GeV by the STAR Collaboration at the BNL Relativistic Heavy Ion Collider}, Phys. Rev. C 105~(1) (2022) 014901.
\newblock \href {http://arxiv.org/abs/2109.00131} {\path{arXiv:2109.00131}}, \href {https://doi.org/10.1103/PhysRevC.105.014901} {\path{doi:10.1103/PhysRevC.105.014901}}.

\bibitem{Voloshin:2010ut}
S.~A. Voloshin, {Testing the Chiral Magnetic Effect with Central U+U collisions}, Phys. Rev. Lett. 105 (2010) 172301.
\newblock \href {http://arxiv.org/abs/1006.1020} {\path{arXiv:1006.1020}}, \href {https://doi.org/10.1103/PhysRevLett.105.172301} {\path{doi:10.1103/PhysRevLett.105.172301}}.

\bibitem{Deng:2016knn}
W.-T. Deng, X.-G. Huang, G.-L. Ma, G.~Wang, {Test the chiral magnetic effect with isobaric collisions}, Phys. Rev. C 94 (2016) 041901.
\newblock \href {http://arxiv.org/abs/1607.04697} {\path{arXiv:1607.04697}}, \href {https://doi.org/10.1103/PhysRevC.94.041901} {\path{doi:10.1103/PhysRevC.94.041901}}.

\bibitem{Anderson:2003ur}
M.~Anderson, et~al., {The Star time projection chamber: A Unique tool for studying high multiplicity events at RHIC}, Nucl. Instrum. Meth. A 499 (2003) 659--678.
\newblock \href {http://arxiv.org/abs/nucl-ex/0301015} {\path{arXiv:nucl-ex/0301015}}, \href {https://doi.org/10.1016/S0168-9002(02)01964-2} {\path{doi:10.1016/S0168-9002(02)01964-2}}.

\bibitem{Llope:2012zz}
W.~J. Llope, {Multigap RPCs in the STAR experiment at RHIC}, Nucl. Instrum. Meth. A 661 (2012) S110--S113.
\newblock \href {https://doi.org/10.1016/j.nima.2010.07.086} {\path{doi:10.1016/j.nima.2010.07.086}}.

\bibitem{Adler:2001fq}
C.~Adler, H.~Strobele, A.~Denisov, E.~Garcia, M.~Murray, S.~White, {The RHIC zero-degree calorimeters}, Nucl. Instrum. Meth. A 461 (2001) 337--340.
\newblock \href {https://doi.org/10.1016/S0168-9002(00)01238-9} {\path{doi:10.1016/S0168-9002(00)01238-9}}.

\bibitem{Llope:2014nva}
W.~J. Llope, et~al., {The STAR Vertex Position Detector}, Nucl. Instrum. Meth. A 759 (2014) 23--28.
\newblock \href {http://arxiv.org/abs/1403.6855} {\path{arXiv:1403.6855}}, \href {https://doi.org/10.1016/j.nima.2014.04.080} {\path{doi:10.1016/j.nima.2014.04.080}}.

\bibitem{SMD}
{STAR Techincal Note SN0448 (2003), Proposed Addition of a Shower Max Detector to the STAR Zero Degree Calorimeters}, \href{https://drupal.star.bnl.gov/STAR/starnotes/public/sn0448}{https://drupal.star.bnl.gov/STAR/starnotes/public/sn0448}.

\bibitem{Miller:2007ri}
M.~L. Miller, K.~Reygers, S.~J. Sanders, P.~Steinberg, {Glauber modeling in high energy nuclear collisions}, Ann. Rev. Nucl. Part. Sci. 57 (2007) 205--243.
\newblock \href {http://arxiv.org/abs/nucl-ex/0701025} {\path{arXiv:nucl-ex/0701025}}, \href {https://doi.org/10.1146/annurev.nucl.57.090506.123020} {\path{doi:10.1146/annurev.nucl.57.090506.123020}}.

\bibitem{Gorbunov:2013yvt}
S.~Gorbunov, {On-line reconstruction algorithms for the CBM and ALICE experiments}, Ph.D. thesis, Goethe U., Frankfurt (main), Frankfurt U. (2013).

\bibitem{Zyzak:2016exl}
M.~Zyzak, {Online selection of short-lived particles on many-core computer architectures in the CBM experiment at FAIR}, Ph.D. thesis, Goethe U., Frankfurt (main), Frankfurt U. (2016).

\bibitem{Kisel:2018nvd}
I.~Kisel, {Event Topology Reconstruction in the CBM Experiment}, J. Phys. Conf. Ser. 1070~(1) (2018) 012015.
\newblock \href {https://doi.org/10.1088/1742-6596/1070/1/012015} {\path{doi:10.1088/1742-6596/1070/1/012015}}.

\bibitem{Voloshin:2016ppr}
S.~A. Voloshin, T.~Niida, {Ultra-relativistic nuclear collisions: Direction of spectator flow}, Phys. Rev. C 94 (2016) 021901(R).
\newblock \href {http://arxiv.org/abs/1604.04597} {\path{arXiv:1604.04597}}, \href {https://doi.org/10.1103/PhysRevC.94.021901} {\path{doi:10.1103/PhysRevC.94.021901}}.

\bibitem{Poskanzer:1998yz}
A.~M. Poskanzer, S.~A. Voloshin, {Methods for analyzing anisotropic flow in relativistic nuclear collisions}, Phys. Rev. C 58 (1998) 1671--1678.
\newblock \href {http://arxiv.org/abs/nucl-ex/9805001} {\path{arXiv:nucl-ex/9805001}}, \href {https://doi.org/10.1103/PhysRevC.58.1671} {\path{doi:10.1103/PhysRevC.58.1671}}.

\bibitem{ParticleDataGroup:2022pth}
R.~L. Workman, et~al., {Review of Particle Physics}, PTEP 2022 (2022) 083C01.
\newblock \href {https://doi.org/10.1093/ptep/ptac097} {\path{doi:10.1093/ptep/ptac097}}.

\bibitem{STAR:2013ayu}
L.~Adamczyk, et~al., {Elliptic flow of identified hadrons in Au+Au collisions at $\sqrt{s_{NN}}=$ 7.7-62.4 GeV}, Phys. Rev. C 88 (2013) 014902.
\newblock \href {http://arxiv.org/abs/1301.2348} {\path{arXiv:1301.2348}}, \href {https://doi.org/10.1103/PhysRevC.88.014902} {\path{doi:10.1103/PhysRevC.88.014902}}.

\bibitem{Borghini:2004ra}
N.~Borghini, J.~Y. Ollitrault, {Azimuthally sensitive correlations in nucleus-nucleus collisions}, Phys. Rev. C 70 (2004) 064905.
\newblock \href {http://arxiv.org/abs/nucl-th/0407041} {\path{arXiv:nucl-th/0407041}}, \href {https://doi.org/10.1103/PhysRevC.70.064905} {\path{doi:10.1103/PhysRevC.70.064905}}.

\bibitem{Karpenko:2016jyx}
I.~Karpenko, F.~Becattini, {Study of $\Lambda $ polarization in relativistic nuclear collisions at $\sqrt{s_\mathrm {NN}}=7.7$ \textendash{}200 GeV}, Eur. Phys. J. C 77~(4) (2017) 213.
\newblock \href {http://arxiv.org/abs/1610.04717} {\path{arXiv:1610.04717}}, \href {https://doi.org/10.1140/epjc/s10052-017-4765-1} {\path{doi:10.1140/epjc/s10052-017-4765-1}}.

\bibitem{Xia:2019fjf}
X.-L. Xia, H.~Li, X.-G. Huang, H.~Z. Huang, {Feed-down effect on \ensuremath{\Lambda} spin polarization}, Phys. Rev. C 100~(1) (2019) 014913.
\newblock \href {http://arxiv.org/abs/1905.03120} {\path{arXiv:1905.03120}}, \href {https://doi.org/10.1103/PhysRevC.100.014913} {\path{doi:10.1103/PhysRevC.100.014913}}.

\bibitem{Becattini:2019ntv}
F.~Becattini, G.~Cao, E.~Speranza, {Polarization transfer in hyperon decays and its effect in relativistic nuclear collisions}, Eur. Phys. J. C 79~(9) (2019) 741.
\newblock \href {http://arxiv.org/abs/1905.03123} {\path{arXiv:1905.03123}}, \href {https://doi.org/10.1140/epjc/s10052-019-7213-6} {\path{doi:10.1140/epjc/s10052-019-7213-6}}.

\bibitem{Li:2021zwq}
H.~Li, X.-L. Xia, X.-G. Huang, H.~Z. Huang, {Global spin polarization of multistrange hyperons and feed-down effect in heavy-ion collisions}, Phys. Lett. B 827 (2022) 136971.
\newblock \href {http://arxiv.org/abs/2106.09443} {\path{arXiv:2106.09443}}, \href {https://doi.org/10.1016/j.physletb.2022.136971} {\path{doi:10.1016/j.physletb.2022.136971}}.

\bibitem{Fang:2016vpj}
R.-H. Fang, L.-G. Pang, Q.~Wang, X.-N. Wang, {Polarization of massive fermions in a vortical fluid}, Phys. Rev. C 94~(2) (2016) 024904.
\newblock \href {http://arxiv.org/abs/1604.04036} {\path{arXiv:1604.04036}}, \href {https://doi.org/10.1103/PhysRevC.94.024904} {\path{doi:10.1103/PhysRevC.94.024904}}.

\bibitem{Csernai:2018yok}
L.~P. Csernai, J.~I. Kapusta, T.~Welle, {$\Lambda$ and $\bar{\Lambda}$ spin interaction with meson fields generated by the baryon current in high energy nuclear collisions}, Phys. Rev. C 99~(2) (2019) 021901.
\newblock \href {http://arxiv.org/abs/1807.11521} {\path{arXiv:1807.11521}}, \href {https://doi.org/10.1103/PhysRevC.99.021901} {\path{doi:10.1103/PhysRevC.99.021901}}.

\bibitem{Vitiuk:2019rfv}
O.~Vitiuk, L.~V. Bravina, E.~E. Zabrodin, {Is different $\Lambda$ and $\bar \Lambda$ polarization caused by different spatio-temporal freeze-out picture?}, Phys. Lett. B 803 (2020) 135298.
\newblock \href {http://arxiv.org/abs/1910.06292} {\path{arXiv:1910.06292}}, \href {https://doi.org/10.1016/j.physletb.2020.135298} {\path{doi:10.1016/j.physletb.2020.135298}}.

\bibitem{Fu:2021pok}
B.~Fu, S.~Y.~F. Liu, L.~Pang, H.~Song, Y.~Yin, {Shear-Induced Spin Polarization in Heavy-Ion Collisions}, Phys. Rev. Lett. 127~(14) (2021) 142301, private communication.
\newblock \href {http://arxiv.org/abs/2103.10403} {\path{arXiv:2103.10403}}, \href {https://doi.org/10.1103/PhysRevLett.127.142301} {\path{doi:10.1103/PhysRevLett.127.142301}}.

\bibitem{Becattini:2021iol}
F.~Becattini, M.~Buzzegoli, G.~Inghirami, I.~Karpenko, A.~Palermo, {Local Polarization and Isothermal Local Equilibrium in Relativistic Heavy Ion Collisions}, Phys. Rev. Lett. 127~(27) (2021) 272302.
\newblock \href {http://arxiv.org/abs/2103.14621} {\path{arXiv:2103.14621}}, \href {https://doi.org/10.1103/PhysRevLett.127.272302} {\path{doi:10.1103/PhysRevLett.127.272302}}.

\bibitem{Becattini:2015ska}
F.~Becattini, G.~Inghirami, V.~Rolando, A.~Beraudo, L.~Del~Zanna, A.~De~Pace, M.~Nardi, G.~Pagliara, V.~Chandra, {A study of vorticity formation in high energy nuclear collisions}, Eur. Phys. J. C 75~(9) (2015) 406, [Erratum: Eur.Phys.J.C 78, 354 (2018)].
\newblock \href {http://arxiv.org/abs/1501.04468} {\path{arXiv:1501.04468}}, \href {https://doi.org/10.1140/epjc/s10052-015-3624-1} {\path{doi:10.1140/epjc/s10052-015-3624-1}}.

\bibitem{Russo:2016ueu}
J.~G. Russo, M.~Tierz, {Quantum phase transition in many-flavor supersymmetric QED$_{3}$}, Phys. Rev. D 95~(3) (2017) 031901.
\newblock \href {http://arxiv.org/abs/1610.08527} {\path{arXiv:1610.08527}}, \href {https://doi.org/10.1103/PhysRevD.95.031901} {\path{doi:10.1103/PhysRevD.95.031901}}.

\bibitem{Fu:2020oxj}
B.~Fu, K.~Xu, X.-G. Huang, H.~Song, {Hydrodynamic study of hyperon spin polarization in relativistic heavy ion collisions}, Phys. Rev. C 103~(2) (2021) 024903.
\newblock \href {http://arxiv.org/abs/2011.03740} {\path{arXiv:2011.03740}}, \href {https://doi.org/10.1103/PhysRevC.103.024903} {\path{doi:10.1103/PhysRevC.103.024903}}.

\bibitem{Yi:2021ryh}
C.~Yi, S.~Pu, D.-L. Yang, {Reexamination of local spin polarization beyond global equilibrium in relativistic heavy ion collisions}, Phys. Rev. C 104~(6) (2021) 064901.
\newblock \href {http://arxiv.org/abs/2106.00238} {\path{arXiv:2106.00238}}, \href {https://doi.org/10.1103/PhysRevC.104.064901} {\path{doi:10.1103/PhysRevC.104.064901}}.

\bibitem{strangev2}
STAR, Elliptic flow of strange and multi-strange hadrons in isobar collisions at $\sqrt{s_{NN}}$ = 200 GeV at RHIC, to be submitted.

\bibitem{Palermo:2024tza}
A.~Palermo, E.~Grossi, I.~Karpenko, F.~Becattini, {$\Lambda $ polarization in very high energy heavy ion collisions as a probe of the quark\textendash{}gluon plasma formation and properties}, Eur. Phys. J. C 84~(9) (2024) 920.
\newblock \href {http://arxiv.org/abs/2404.14295} {\path{arXiv:2404.14295}}, \href {https://doi.org/10.1140/epjc/s10052-024-13229-z} {\path{doi:10.1140/epjc/s10052-024-13229-z}}.

\end{thebibliography}
